\documentclass[aps,twocolumn,footinbib,floatfix,
superscriptaddress,
longbibliography,
preprintnumbers]{revtex4-1}

\usepackage{amsmath,amssymb,bm,multirow,float,array,bbold,subfigure}
\usepackage{graphicx,color}
\usepackage[usenames,dvipsnames]{xcolor}
\usepackage{dsfont} 
\usepackage[colorlinks,bookmarksopen,bookmarksnumbered,pdfstartview=FitH,citecolor=SBred, linkcolor=SBred,urlcolor=SBred]{hyperref}
\usepackage{graphicx}
\usepackage{enumitem}

\definecolor{SBred}{rgb}{0.6471, 0.1098, 0.1882}

\def\a23{\alpha_{23}}

\newcommand{\GeV}{\,{\rm GeV}}

\newcommand{\be}{\begin{equation}}
\newcommand{\ee}{\end{equation}}

\newcommand{\eq}[1]{\begin{equation} #1 \end{equation}}
\newcommand{\eqa}[1]{\begin{eqnarray} #1 \end{eqnarray}}

\newcommand{\ord}{\mathcal{O}}

\newcommand{\A}{{\cal A}}
\newcommand{\F}{{\cal F}}
\renewcommand{\H}{{\cal H}}
\newcommand{\N}{{\cal N}}

\newcommand{\K}{{\cal K}}

\renewcommand{\O}{{\cal O}}
\newcommand{\av}[1]{\langle #1 \rangle}
\newcommand{\ite}{\noindent $\blacktriangleright$ }

\newcommand{\Ref}[1]{Ref.~\cite{#1}}
\newcommand{\Fig}[1]{Fig.~\ref{#1}}
\newcommand{\Eq}[1]{Eq.~(\ref{#1})}
\newcommand{\Tab}[1]{Table~\ref{#1}}
\newcommand{\Sec}[1]{Section~\ref{#1}}

\def\beqn#1{\begin{equation}\label{#1}}
\def\eeqn{\end{equation}}
\def\beqa#1{\begin{eqnarray}\label{#1}}
\def\eeqa{\end{eqnarray}}

\begin{document}

\preprint{EOS-2017-01}
\preprint{MIT-CTP 4918}
\preprint{TUM-HEP-1087/17}
\preprint{ZU-TH 17/17}

\title{Long-distance effects in $B\to K^*\ell\ell$ from Analyticity}

\author{Christoph Bobeth} 
\affiliation{
Physik Department, TU M\"unchen, James-Franck-Stra{\ss}e 1, D-85748 Garching, Germany
}
\affiliation{
Excellence Cluster Universe, Technische Universit\"at M\"unchen, Boltzmannstr. 2, D-85748 Garching, Germany
}

\author{Marcin Chrzaszcz}
\affiliation{
Physik-Institut, Universit\"at Z\"urich, Winterthurer Strasse 190, 8057 Z\"urich, Switzerland
}
\affiliation{
H.Niewodniczanski, Institute of Nuclear Physics Polish Academy of Sciences, ul. Radzikowskiego 152, 31-342 Krakow, Poland
}

\author{Danny van Dyk}
\affiliation{
Physik-Institut, Universit\"at Z\"urich, Winterthurer Strasse 190, 8057 Z\"urich, Switzerland
}

\author{Javier Virto}
\affiliation{
Physik Department, TU M\"unchen, James-Franck-Stra{\ss}e 1, D-85748 Garching, Germany
}
\affiliation{
Center for Theoretical Physics, Massachusetts Institute of Technology, Cambridge, MA 02139, USA
}


\begin{abstract}
\vspace{5mm}

We discuss a novel approach to systematically determine the long-distance contribution to $B\to K^*\ell\ell$
decays in the kinematic region where the dilepton invariant mass is below the open charm threshold. This approach provides the most consistent and reliable
determination to date and can be used to compute Standard Model predictions for all observables
of interest, including the kinematic region where the dilepton invariant mass lies
between the $J/\psi$ and the $\psi(2S)$ resonances.
We illustrate the power of our results by performing a New Physics fit to the
Wilson coefficient $C_9$.
This approach is systematically improvable from theoretical and experimental sides,
and applies to other decay modes of the type $B\to V\ell\ell$, $B\to P\ell\ell$ and $B\to V\gamma$.

\vspace{3mm}
\end{abstract}

\maketitle

\section{Introduction}

$B\to K^*\ell\ell$ decays are sensitive to modified short-distance physics from
sources beyond the Standard Model (SM), and a great deal of experimental 
and theoretical work has been devoted to extract short-distance information
from them. However, long-distance physics within the SM also contributes
significantly to the decay, and its effects are very difficult to assess reliably
from first principles.
On the other hand, tighter experimental constraints from increasingly
precise measurements of $b\to s$ processes have significantly limited the size of allowed New Physics (NP) effects in $B\to K^*\ell\ell$, which are now comparable 
to current SM uncertainties.
Thus, our inability to reliably constrain these long-distance contributions to acceptable levels stands in the way of 
obtaining unambiguous information on physics beyond the SM.

The $B\to K^*\ell\ell$ decay is conveniently described by the $K^*$ transversity
amplitudes ($\lambda = \perp, \parallel, 0$)
\eqa{
\A_\lambda^{L,R} &=& \N_\lambda\ \bigg\{ 
(C_9 \mp C_{10}) \F_\lambda(q^2) \\
&&+\frac{2m_b M_B}{q^2} \bigg[ C_7 \F_\lambda^{T}(q^2) - 16\pi^2 \frac{M_B}{m_b} \H_\lambda(q^2) \bigg]
\bigg\}\nonumber 
}
where $C_{7,9,10}$ are short-distance Wilson coefficients, and $\N_\lambda$ are normalization factors.
The non-trivial matter from the theory point of
view is the determination of the ``local'' and ``non-local'' long-distance effects encoded in
the functions $\F_\lambda^{(T)}(q^2)$ and $\H_\lambda(q^2)$, respectively, which
depend on the dilepton invariant mass squared $q^2$.

The functions $\F_\lambda^{(T)}(q^2)$ are form factors, which
can be calculated by means of Light-Cone Sum Rules (LCSRs) at low $q^2$~($\lesssim 10\,\GeV^2$)~\cite{Ball:1998kk,Khodjamirian:2006st},
or by numerical simulations (Lattice QCD) at large $q^2$~($\gtrsim 15\,\GeV^2$)~\cite{Becirevic:2006nm,Horgan:2013hoa}.
Both methods agree reasonably well when extrapolated~\cite{Descotes-Genon:2013vna,Straub:2015ica},
and there are good prospects for improvement~\cite{Hansen:2012tf,Briceno:2014uqa,Cheng:2017smj,Boer:2016iez,Wang:2015vgv}.
The form factors are not the main focus of this work.

Here we focus on the functions $\H_\lambda(q^2)$, which are related to the
contribution from 4-quark and chromomagnetic operators in the Weak Effective
Hamiltonian, and emerge from the ``non-local'' matrix element
\eq{
\eta^*_\alpha \, \H^{\alpha\mu} = i \int d^4 x\ e^{i q\cdot x}\,
\av{\bar K^*(k,\eta)|\K^\mu(x,0) | \bar B(p)}\ ,
\label{eq:Correlator}
}
where $p=q+k$, $\eta$ is the polarization vector of the $K^*$, and $\K(x,y)$ is a bi-local operator.
The most relevant contribution to this matrix element in the SM arises from the current-current operators $\O_{1,2}$,
since they come with large Wilson coefficients.
In this letter we consider only this contribution -- the so-called ``charm-loop effect'' -- for which the
object $\K^\mu(x,y)$ is given by:
\eq{
\K^\mu(x,y) =  T\{ j_{\rm em}^\mu(x), C_1 \O_1(y) + C_2 \O_2(y) \}
}
with $ j_{\rm em}^\mu(x) = \sum_q Q_q\,\bar q(x) \gamma^\mu q(x)$ the electromagnetic
current. The scalar functions $\H_\lambda(q^2)$ are given by the Lorentz decomposition:
\eq{
\H^{\alpha\mu}(q,k) = M_B^2 \, \big[
S^{\alpha\mu}_\perp \,\H_\perp - S^{\alpha\mu}_\|\, \H_\| - S^{\alpha\mu}_0\, \H_0
\big]
\label{LorDecH}
}
where $S_\lambda^{\alpha\mu}$ are a set of structures given in the appendix.

In the heavy $b$-quark limit and for very small $q^2$, the functions $\H_\lambda(q^2)$
factorize into non-perturbative form factors and light-cone distribution amplitudes,
up to perturbatively calculable ``hard'' functions~\cite{Beneke:2001at}.
However this perturbative expansion breaks down when $q^2$ approaches $4 m_c^2$,
leading to questionable predictions for $q^2\gtrsim 6\GeV^2$.
The integral in \Eq{eq:Correlator} is in fact dominated by the region
$x^2\lesssim (2m_c - \sqrt{q^2})^{-2}$~\cite{Khodjamirian:2010vf},
so for $q^2\ll 4m_c^2$ one may expand the operator
$\K^\mu(x,0)$ around $x^2 = 0$ (a light-cone operator-product expansion, or LCOPE).
This leads to an expansion of \Eq{eq:Correlator} in powers of $(2m_c - \sqrt{q^2})^{-1}$,
with matrix elements of operators that are non-local only along the light cone.
This theory framework has been worked out up to NLO in $\alpha_s$~\cite{Beneke:2001at,Asatryan:2001zw}
and including subleading terms in the LCOPE~\cite{Khodjamirian:2010vf},
and can be safely applied for $q^2\ll 4m_c^2$ (preferably at $q^2<0$).
However, reliable predictions for larger values of $q^2$ remain a challenge.

In this letter we consider a consistent, model-independent and systematically-improvable
approach to determine the long-distance contributions $\H_\lambda(q^2)$ to $B\to K^*\ell\ell$
in the region $q^2 \lesssim 14\GeV^2$.
It provides genuine SM predictions even in the presence of NP in semileptonic operators.
In addition, this approach provides access to the inter-resonance region $10\GeV^2 \lesssim q^2 \lesssim 13\GeV^2$. The idea is the following: We determine the analytic properties
of the functions $\H_\lambda(q^2)$ in the complex plane, and use this information to
write down general and model-independent parametrizations. We then use two
pieces of information to constrain the parametrized functions: data
on $B\to K^* J/\psi$ and $B\to K^* \psi(2S)$, which is independent of NP in
semileptonic operators; and theory at $q^2<0$, where it is reliable.
This method, which builds upon
Refs.~\cite{Khodjamirian:2010vf,Khodjamirian:2012rm}, gives the most reliable
and consistent \textit{a-priori} determination of the functions
$\H_\lambda(q^2)$ to date. We use these results to compute
SM predictions (assuming no NP in $\O_{1,2}$), and to perform a NP fit to $C_9$.
All our numerical computations are performed with the help of
EOS~\cite{EOS:web}, which has been modified for this
purpose~\cite{EOS:release}.

\section{Analytic Structure and Parametrization}

It is a standard assumption in quantum field theory that the only analytic singularities
of a correlation function -- as a complex function of all its complexified kinematic invariants --
are those required by unitarity~\cite{Eden:1966dnq}.
Unitarity, in turn, relates analytic singularities with on-shell intermediate states:
poles for one-particle states, and branch cuts for multi-particle states.
Thus, the analytic structure of a correlation function can be learned by analysing its on-shell cuts.

In the case at hand, inspection of the correlation function~(\ref{eq:Correlator}) reveals the following
analytic properties of the scalar functions $\H_\lambda(q^2)$:\\  

\ite On-shell cuts in the variable $q^2$ include:
two poles at $q^2=M_{J/\psi}^2\simeq 9\GeV^2$ and
$q^2=M_{\psi(2S)}^2\simeq 14\GeV^2$ corresponding to
one-particle intermediate states through $B\to K^* \psi_n (\to \ell^+\ell^-)$,
with $\psi_1 = J/\psi$ and $\psi_2 = \psi(2S)$;
a branch cut starting at $q^2=t_+\equiv 4 M_D^2$ corresponding to two-particle
intermediate states through $B\to K^* [\bar D D] (\to \ell^+\ell^-)$, plus
other ``$c\bar c$'' cuts with higher thresholds;
and ``light-hadron'' branch cuts starting at $q^2\simeq 0$ from intermediate states
such as $B\to K^* [3\pi](\to \ell^+\ell^-)$, which include finite-width
effects of $J/\psi$ and $\psi(2S)$. The effects of these ``light-hadron'' cuts are OZI suppressed~\cite{Okubo:1963fa,Zweig:1964jf,Iizuka:1966fk}.
Given the limited precision of current data, we will neglect these OZI suppressed
contributions, as well as the effects of other light hadron resonances.

\bigskip

\ite On-shell cuts in the variable $(q+k)^2$ (the ``forward'' or ``decay'' channel) include branch cuts
from intermediate states such as $B\to \bar D D_s \to K^*\ell^+\ell^-$. The physical point $(q+k)^2=M_B^2$
lies on these cuts, which implies that the functions $\H_\lambda(q^2)$ are complex-valued for all values of $q^2$.
But this imaginary part is not associated with any singularity in the variable $q^2$. Thus, one can write
$\H_\lambda(q^2) = \H_\lambda^{\rm (re)}(q^2) + i\, \H_\lambda^{\rm (im)}(q^2)$, with $\H_\lambda^{\rm (re,im)}(q^2)$
satisfying the analytic properties of the previous point as functions of $q^2$, and obeying the same dispersion relation.\\

These properties can be exploited to write down a general parametrization for
the correlator consistent with unitarity. A convenient way to do so is to
re-express the functions $\H_\lambda(q^2)$ in terms of the ``conformal''
variable $z$:
\eq{
z(q^2) \equiv \frac{\sqrt{t_+ - q^2} - \sqrt{t_+ - t_0}}{\sqrt{t_+ - q^2} + \sqrt{t_+ - t_0}}\,,
}
where $t_+= 4M_D^2$ and $t_0= t_+ - \sqrt{t_+(t_+-M_{\psi(2S)}^2)}$. This transformation maps the $c\bar c$ branch cut
in the $q^2$ plane to the unit circumference $|z|=1$, and the entire first Riemann sheet in the $q^2$ plane to the interior
of the unit circle $|z|<1$. Our choice for $t_0$ implies that within the relevant interval $-7\GeV^2 \leq q^2 \leq M_{\psi(2S)}^2$,
$|z| < 0.52$.

The approach now resembles and is inspired by the $z$-parametrization used for the form factors~\cite{Boyd:1995cf,Bourrely:2008za}.
The functions $\H_\lambda(z) \equiv \H_\lambda(q^2(z))$ are meromorphic in $|z|<1$,
with two simple poles at $z_{J/\psi}\equiv z(M_{J/\psi}^2)\simeq 0.18$ and $z_{\psi(2S)}\equiv z(M_{\psi(2S)}^2)\simeq -0.44$.
Therefore, multiplying by the corresponding zeroes will give an analytic function in $|z|<1$ that can be Taylor-expanded around $z=0$. This expansion should converge reasonably well in the region of interest, where $|z| < 0.52$.
This is the basis of our proposed parametrization.

In order to assure that the leading terms in the expansion will capture most of the features of the function
(thus improving convergence), we use two more pieces of information: First, the correlator inherits all the singularities
of the form factor ({\it e.g.} the $M_{B_s^*}$ pole), and the leading OPE contribution to the correlator is
indeed proportional to the form factor. Therefore it is better to parametrize the ratios $\H_\lambda(q^2)/\F_\lambda(q^2)$ instead.
Second, the poles should not modify the asymptotic behaviour. This is achieved by introducing
appropriate ``Blaschke factors''~\cite{Boyd:1995cf}. All in all, we propose the following parametrization:
\eq{
\H_\lambda(z) =
\frac{1-z\, z^*_{J/\psi}}{z-z_{J/\psi}} \frac{1-z\,z^*_{\psi(2S)}}{z-z_{\psi(2S)}} \hat\H_\lambda(z)\ ,
\label{H}
}
with
\eq{
\hat\H_\lambda(z) = \Big[ \sum_{k=0}^K \alpha_k^{(\lambda)} z^{k} \Big] \F_\lambda(z)\ ,
\label{Hhat}
}
where $\alpha^{(\lambda)}_k$ are complex coefficients, and the expansion is truncated after the term $z^{K}$.
This truncation unavoidably introduces some model dependence.
The maximum value that can be chosen for $K$ will depend on the available set
of experimental measurements and theory inputs.

\begin{table}[b]
\centering
\renewcommand{\arraystretch}{1.5}
\renewcommand{\tabcolsep}{3.1mm}
\begin{tabular}{@{}crrr@{}}
\hline
$k$  &  0\hspace{7mm} & 1\hspace{7mm} & 2\hspace{7mm} \\
\hline
${\rm Re}[\alpha_{k}^{(\perp)}]$ & $-0.06 \pm 0.21$  & $-6.77 \pm 0.27$ & $18.96 \pm 0.59$ \\
${\rm Re}[\alpha_{k}^{(\parallel)}]$ & $-0.35 \pm 0.62$  & $-3.13 \pm 0.41$ & $12.20 \pm 1.34$ \\
${\rm Re}[\alpha_{k}^{(0)}]$ & $0.05 \pm 1.52$  & $17.26 \pm 1.64$ & --  \\
${\rm Im}[\alpha_{k}^{(\perp)}]$ & $-0.21 \pm 2.25$  &  $1.17 \pm 3.58$   & $-0.08 \pm 2.24$ \\
${\rm Im}[\alpha_{k}^{(\parallel)}]$ & $-0.04 \pm 3.67$  & $-2.14 \pm 2.46$   &  $6.03 \pm 2.50$ \\
${\rm Im}[\alpha_{k}^{(0)}]$  & $-0.05 \pm 4.99$  &  $4.29 \pm 3.14$ & --  \\
\hline        
\end{tabular}
\caption{Mean values and standard deviations (in units of $10^{-4}$)
of the prior PDF for the parameters $\alpha_k^{(\lambda)}$.}
\label{alphak}
\end{table}

\section{Experimental constraints}
\label{experimentalconstraints}

According to the LSZ reduction formula \cite{Weinberg:1995mt}, the amplitudes for
the decays $B\to K^* \psi_n$ (with $\psi_1 = J/\psi$ and $\psi_2= \psi(2S)$)
are defined by the residues of the functions $\H_\lambda(q^2)$ on the $\psi_n$
poles:
\eq{
\H_\lambda(q^2 \to M_{\psi_n}^2) \sim
\frac{M_{\psi_n} f^*_{\psi_n} \A^{\psi_n}_\lambda}{M_B^2 (q^2 - M_{\psi_n}^2)} + \cdots\,,
}
where the dots represent regular terms.
Here $\av{0| j_{\rm em}^\mu |\psi_n(q,\varepsilon)} = M_{\psi_n} f^*_{\psi_n} \varepsilon^\mu$, and $\A^{\psi_n}_\lambda$ are the $B\to K^* \psi_n$ transversity amplitudes.
The most precise constraints on these amplitudes can be obtained from
Babar~\cite{Aubert:2004rz,Aubert:2007hz},
Belle~\cite{Itoh:2005ks,Chilikin:2013tch,Chilikin:2014bkk} and
LHCb~\cite{Aaij:2013cma}.

We use the data to produce two sets of five pseudo-observables (three magnitudes and two relative phases on each
resonance):
\eq{
|r_\perp^{\psi_n}|,\,
|r_\|^{\psi_n}|,\,
|r_0^{\psi_n}|,\,
\arg\{r_\perp^{\psi_n} r_{0}^{\psi_n*}\},\,
\arg\{r_\|^{\psi_n} r_{0}^{\psi_n*}\},
\label{pseudoexp}
}
where
\eq{
r_\lambda^{\psi_n} \equiv \operatorname*{Res}_{q^2\to M^2_{\psi_n}} \frac{\H_\lambda(q^2)}{\F_\lambda(q^2)}
\sim
\frac{M_{\psi_n} f^*_{\psi_n} \A^{\psi_n}_\lambda}{M_B^2\, \F_\lambda(M_{\psi_n}^2)}\ .
\label{r_lambda}
}
The numerical values for these pseudo-observables are obtained from the posterior-predictive distributions of a Bayesian
fit. The inputs for this fit and the results are provided for completeness in the appendix.
These pseudo-observables will act as constraints on the parameters of the correlators at $z = 0.18$ and $z = -0.44$.

\section{Theory constraints}
\label{theoryconstraints}

At $q^2<0$ the functions $\H_\lambda$ can be calculated with the current
approaches for the large recoil region. We use QCD-factorization at
next-to-leading order in $\alpha_s$, including the form factor terms and
hard-spectator contributions~\cite{Beneke:2001at,Beneke:2004dp}.  In addition,
we include~\footnote{%
    We thank Yuming Wang for providing us with the results for $B\to K^*\gamma^*$
    in digital form.
} the soft-gluon correction calculated via a LCSR in
Ref.~\cite{Khodjamirian:2010vf}.  For the form factors we use the results from
the LCSR with $B$-meson distribution amplitudes~\cite{Khodjamirian:2006st}, in
order to have a mutually consistent description of form factors and non-local
contributions and benefit from theoretical correlations among both.  In this way
we compute the ratios $\H_\lambda(q^2)/\F_\lambda(q^2)$ at the
points $q^2=\{-7,-5,-3,-1\}\GeV^2$. These ratios are
used as pseudo-observables to constrain the
parameters in~\Eq{H} at $z=\{0.52,0.50,0.48,0.46\}$.
Further details and results are presented for completeness in the appendix.
We emphasize that no theory is used at $q^2 \ge 0$ at all.

\section{SM predictions}
\label{sec:smpreds}

\begin{figure}
\includegraphics[width=9cm]{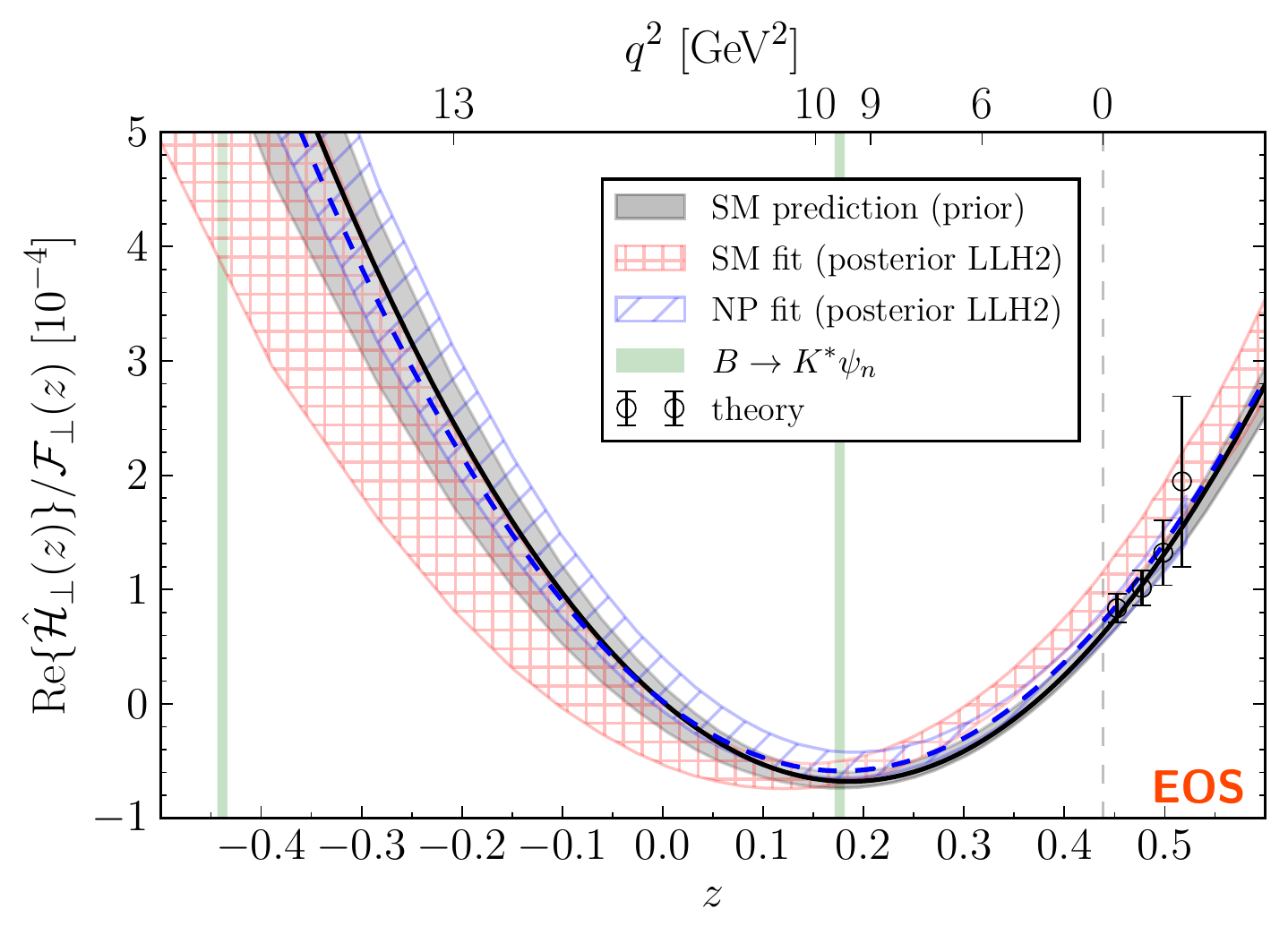}
\caption{Results of the prior and posterior fits for the ratio ${\rm
Re}[{\hat \H}_\perp(z)]/\F_\perp(z)$.  See the text for details.}
\label{fig1}
\end{figure}

We now perform a fit of \Eq{H} to the combined experimental and theoretical
constraints described above in Sections~\ref{experimentalconstraints} and
\ref{theoryconstraints}. 
We find that \Eq{H} with $K=2$ provides an excellent fit to all inputs, with a $p$-value
of $0.91$. All 1D-marginalised posteriors are reasonably symmetric around their
modes. The result of this fit is a set of correlated values
for the complex parameters $\alpha^{(\lambda)}_k$, which are summarized in Table~\ref{alphak}.
These values lead
to a determination of the non-local correlator in \Eq{eq:Correlator} that is
consistent with the $B\to K^*\psi_n$ measurements, the theory calculations
at negative $q^2$, and it is independent of new physics in semileptonic
operators. Thus, unlike \Ref{Ciuchini:2015qxb}, this is a \emph{genuine SM determination}.

The gray band in \Fig{fig1} shows the result of
this ``prior'' fit for the case of the real part of $\H_\perp(q^2)$. Similar
plots for the other correlators are provided in the appendix for completeness.

\begin{figure}
\includegraphics[width=9cm]{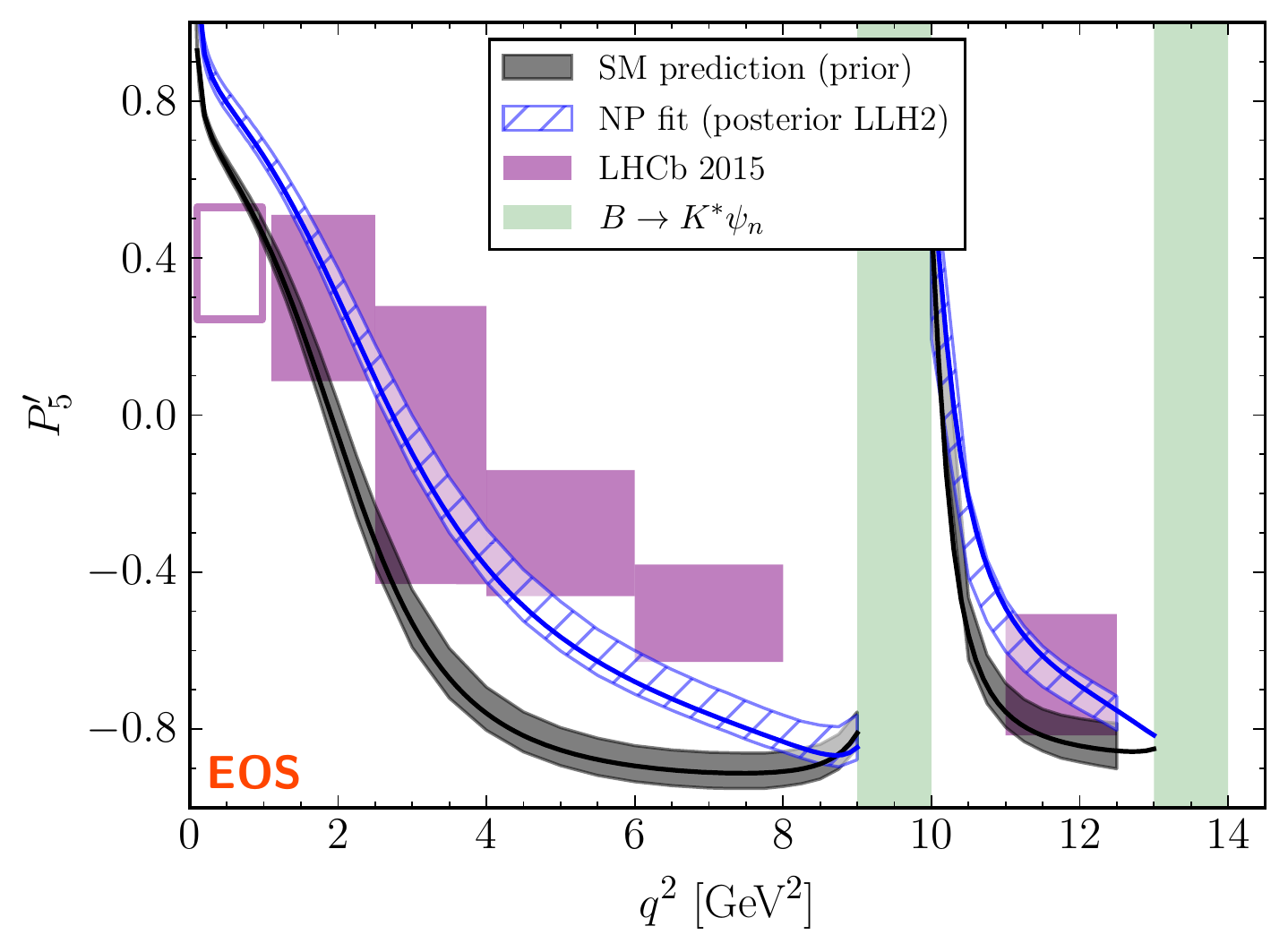}
\caption{Prior and posterior predictions for $P_5'$ within the SM and the NP $C_9$ benchmark,
compared to LHCb data.}
\label{fig2}
\end{figure}

With these results at hand, we can compute SM predictions for all observables of interest within the
range $0\le q^2 \lesssim 14\GeV^2$.
One of them is the angular observable $P'_5$~\cite{DescotesGenon:2012zf}, which is the visible
face of the ``$B\to K^*\mu^+\mu^-$ anomaly''~\cite{Aaij:2013qta}.
Our SM prediction for $P_5'$ is represented by the gray band in \Fig{fig2}. We find
relatively small uncertainties and a clearly apparent tension with LHCb data (represented by purple boxes
in \Fig{fig2}).

Another interesting SM prediction that we obtain from our analysis is:
\eq{
\begin{aligned}
BR(B^0\to K^{*0}\gamma) & = (4.2^{+1.7}_{-1.3}) \cdot 10^{-5}\ ,
\end{aligned}
}
in agreement with the world average \cite{Amhis:2016xyh}.
The larger uncertainties as compared to~\Ref{Paul:2016urs} are due to
our doubling of the form factor uncertainties.
SM predictions for all other observables will be given elsewhere.

\section{New physics analysis}

We now perform a fit to $B\to K^*\mu^+\mu^-$ data using as prior information
the SM predictions derived in \Sec{sec:smpreds}.
We include the branching ratio and the angular observables
$S_i$~\cite{Altmannshofer:2008dz} within the $q^2$ bins in the region $1 \le q^2 \lesssim 14\GeV^2$.
We use the latest LHCb measurements~\cite{Aaij:2015oid,Aaij:2016flj}, and perform different separate fits,
using the results from the maximum-likelihood fit excluding (LLH) and including (LLH2) the inter-resonance bin,
or using the results from the method of moments~\cite{Beaujean:2015xea} (MOM and MOM2),
and both including (NP fit) and not including (SM fit) a floating NP contribution to $C_9$.

The fits provide posterior distributions for the correlator, for $B\to K^*\mu^+\mu^-$ and $B\to K^*\gamma$
observables, and for $C_9$. We first discuss some illustrative results of the LLH2 fit.
The posteriors for the real part of $\H_\perp(q^2)$ are shown in \Fig{fig1}, both for the SM and the NP fits.
In this case it is reassuring that both are consistent within errors with the result of the prior fit,
indicating that modifying the long-distance contribution does not lead to improvement in the SM fit,
and so the long-distance contribution is not likely to mimic a NP contribution.

The posterior NP prediction for $P_5'$ (corresponding to the LLH2 fit) is shown
in \Fig{fig2}, exhibiting a much better agreement with the experimental
measurements than the SM (prior) prediction.

\begin{figure}
\includegraphics[width=\columnwidth]{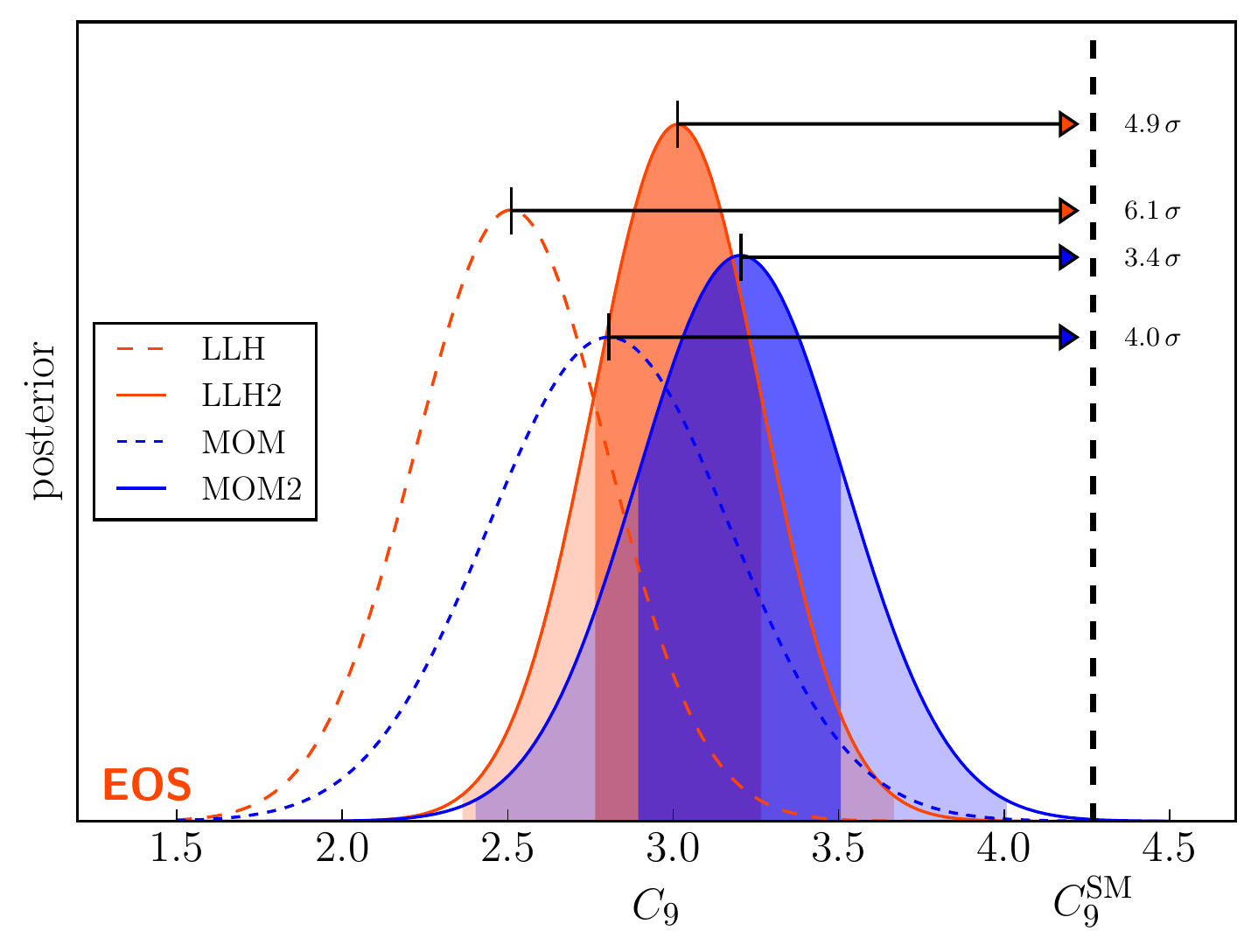}
\caption{Posterior distributions for $C_9$ from the NP fits
and their respective pulls.
Dark and light shaded regions correspond to 68\% and 99\%
probability.
}
\label{fig3}
\end{figure}

The main conclusion of the fits is the following. The SM fits are relatively inefficient
in comparison with the NP fits, with posterior odds \cite{Beaujean:2013soa} ranging from $\sim 2.7$ to $\sim 10$
(on the log scale) in favor of the NP hypothesis.
The one-dimensional marginalized posteriors yield:
\allowdisplaybreaks
\begin{align}
    \text{(LLH)} : C_9 & = 2.51 \pm 0.29\,,\\
    \text{(LLH2)}: C_9 & = 3.01 \pm 0.25\,,\\
    \text{(MOM)} : C_9 & = 2.81 \pm 0.37\,,\\
    \text{(MOM2)}: C_9 & = 3.20 \pm 0.31\,.
\end{align}
The corresponding pulls with respect to the SM point $C_9^{\rm SM}(\mu = 4.2\,\GeV) = 4.27$
range from $3.4$ to $6.1$ standard deviations, and are illustrated in \Fig{fig3}.
These results, from a fit to $B\to K^*\mu^+\mu^-$ data only, are in qualitative agreement
with \emph{global} fits~\cite{%
Descotes-Genon:2013wba,%
Altmannshofer:2013foa,%
Beaujean:2013soa,%
Altmannshofer:2017fio,%
Capdevila:2017bsm,%
Geng:2017svp,%
Hurth:2017hxg%
},
but rely on a more fundamented theory treatment.

\section{Conclusions}

Analyticity provides strong constraints on the hadronic contribution to $B\to K^*\ell\ell$ observables,
and fixes the $q^2$ dependence up to a polynomial, which under some circumstances is an expansion in
a small kinematical parameter. In this letter we have exploited this idea to propose a systematic
approach to determine the non-local contributions, which at this time are the main source of theory
uncertainty. This approach is systematically improvable with more precise data on $B\to K^*\psi_n$
and/or more precise theory calculations at negative $q^2$. In addition, this approach allows access to
the inter-resonance region, which provides valuable information on short-distance physics.
We have focused on $B\to K^*\ell\ell$, but the approach applies to any other $B\to M\ell\ell$
modes such as $B\to \lbrace K,\pi,\rho\rbrace\ell\ell$ and $B_s\to \phi\ell\ell$.

We have performed a numerical analysis implementing this idea, and conclude that significantly improved
theory predictions can be obtained, leading to a more precise and robust interpretation of experimental data
and an improved sensitivity to short-distance physics.
We thus believe that this approach will become very useful in future analyses of exclusive $b\to s$ and $b\to d$ transitions.

\bigskip

\noindent  
{\it Acknowledgements}\\[2mm]
\linespread{1}\selectfont
{\small
We thank Wolfgang Altmannshofer, Kirill Chilikin, Gilberto Colangelo, Christoph Hanhart, Bastian Kubis, Alexander Khodjamirian, Thomas Mannel, Joaquim Matias, Mikolaj Misiak, Federico Mescia, Jacobo Ruiz de Elvira, David Straub,
Lewis Tunstall, Yuming Wang and Roman Zwicky for useful interactions and discussions.
CB is supported in part by the DFG SFB/TR~110 ``Symmetries and the Emergence of Structure in QCD''.
CB and DvD acknowledge support from the Munich Institute for Astro- and Particle Physics (MIAPP)
of the DFG cluster of excellence ``Origin and Structure of the Universe''.
MC is grateful for support of the Polish National Science Center
under the ”Sonata” grant: UMO-2015/17/D/ST2/03532.
DvD is supported in part by the Swiss National Science Foundation (SNF) under contract 200021-159720.
JV acknowledges funding from the Swiss National Science Foundation, from
Explora project FPA2014-61478-EXP, and from the European Union's Horizon 2020
research and innovation programme under the Marie Sklodowska-Curie grant
agreement No 700525 `NIOBE'.
This research was supported in part by PL-Grid Infrastructure.
}

\appendix

\section{Supplemental details and results}

The effective Lagrangian that governs $b\to s$ transitions contains the following terms relevant for our analysis:
\eq{
\mathcal{L}_\text{eff} = \frac{4 G_F}{\sqrt{2}} V_{tb}^{} V_{ts}^*  \sum_{i=1,2,7,9,10} C_i(\mu)\,\O_i(\mu) + \cdots
}
with the current-current operators defined as
\eqa{
    \O_1 & =& \left[\bar{s} \gamma^\mu P_L T^A c\right] \, \left[\bar{c} \gamma_\mu P_L T^A b\right]\,,\\[2mm]
    \O_2 & =& \left[\bar{s} \gamma^\mu P_L c\right]     \, \left[\bar{c} \gamma_\mu P_L b\right]\,,
}
such that $C_2(\mu = M_W) = 1 + \ord(\alpha_s)$, with $P_{L(R)} = (1 \mp \gamma_5)/2$
and $T^A$ the generators of SU(3), and the dipole and semileptonic operators given by
\eqa{
    \O_7    & = & \frac{e\,m_b}{(4\pi)^2} \left[\bar{s} \sigma^{\mu\nu} P_R b\right]\,F_{\mu\nu}\,, \\
    \O_9    & = & \frac{\alpha_e}{4\pi} \left[\bar{s} \gamma^\mu P_L b\right]\,\left[\bar\ell \gamma_\mu \ell\right]\,, \\
    \O_{10} & = & \frac{\alpha_e}{4\pi} \left[\bar{s} \gamma^\mu P_L b\right]\,\left[\bar\ell \gamma_\mu \gamma_5 \ell\right]\,.
}
In the SM, the values of the Wilson coefficients are:
\eq{
\begin{aligned}
C_1^{\rm SM}(m_b) &= -0.3\ , & C_2^{\rm SM}(m_b) &= 1.0\ ,  \\
C_7^{\rm SM}(m_b) &= -0.3\ , & & \\
C_9^{\rm SM}(m_b) &= 4.3\ ,  & C_{10}^{\rm SM}(m_b) &= -4.2\ .
\end{aligned}
}

The Lorentz decomposition for the form factors that we use in this analysis
is given by
\allowdisplaybreaks
\eqa{
\av{ \bar s\gamma^\mu b } &=&
\eta^*_\alpha \, S^{\alpha\mu}_\bot {\cal F}_\bot\ ,
\nonumber \\[2mm]
\av{\bar s\gamma^\mu\gamma_5 b } &=&
\eta^*_\alpha \, (S^{\alpha\mu}_\| {\cal F}_\| + S^{\alpha\mu}_0 {\cal F}_0 + S^{\alpha\mu}_t {\cal F}_t)\ ,
\nonumber \\[2mm]
\av{\bar s\sigma^{\mu\nu}q_\nu b } &=&
i M_B\ \eta^*_\alpha \, S^{\alpha\mu}_\bot {\cal F}_\bot^T\ ,
\label{formfactors}\\[2mm]
\av{ \bar s\sigma^{\mu\nu}q_\nu\gamma_5 b } &=&
-i M_B\ \eta^*_\alpha \big(S^{\alpha \mu}_\| {\cal F}_\|^T + S^{\alpha\mu}_0 {\cal F}_0^T \big)\ ,
\nonumber 
}
denoting $\av{\Gamma}\equiv \av{\bar K^*(k,\eta) | \Gamma | \bar B (q+k)}$.
The non-local correlator $\H^{\alpha\mu}$ is decomposed analogously according to \Eq{LorDecH}:
\eq{
\H^{\alpha\mu}= M_B^2 \, \big[
S^{\alpha\mu}_\perp \,\H_\perp - S^{\alpha\mu}_\|\, \H_\| - S^{\alpha\mu}_0\, \H_0
\big]\ .
\nonumber
}
The Lorentz structures $S_\lambda^{\alpha\mu}$ are given by: \allowdisplaybreaks
\eqa{
S^{\alpha\mu}_\perp &=& \frac{\sqrt{2} M_B}{\sqrt{\lambda}} \varepsilon^{\alpha\mu k q}\ , \nonumber\\[2mm]
S^{\alpha\mu}_\| &=& \frac{i M_B}{\sqrt{2} \lambda} \Big[ \lambda g^{\alpha\mu} + 4 M_{K^*}^2 q^\alpha q^\mu - 4 (q\cdot k)\, q^\alpha k^\mu \Big]\ , \nonumber \\[2mm]
S^{\alpha\mu}_0 &=& -\frac{i\, 4 M_{K^*} (M_B + M_{K^*})} {\lambda\sqrt{q^2}} \Big[(q\cdot k)\, q^\alpha q^\mu - q^2\, q^\alpha k^\mu  \Big]\ , \nonumber \\[2mm]
S^{\alpha\mu}_t &=& \frac{i\, 2 M_{K^*}}{q^2} q^\alpha q^\mu\ .
}
\begin{table}
    \centering
    \renewcommand{\arraystretch}{1.4}
    \renewcommand{\tabcolsep}{4.1mm}
\begin{tabular}{@{}ccccc@{}}
        \hline
        Parameter                                   & Prior ($68\%$ gaussian)     \\
        \hline 
        $\lambda$                                   & $0.225$   $\pm$ $0.006$   \\
        $A$                                         & $0.829 $  $\pm$ $0.012$   \\
        $\bar{\rho}$                                & $0.132 $  $\pm$ $0.018$   \\
        $\bar{\eta}$                                & $0.348 $  $\pm$ $0.012$   \\
        \hline
    \end{tabular}
    \caption{%
        Uncorrelated priors for the CKM parameters in our analysis, taken from the tree-level-only fit in \Ref{Bona:2006ah}.
    }
    \label{CKM}
\end{table}

\begin{table}
    \centering
    \renewcommand{\arraystretch}{1.6}
    \renewcommand{\tabcolsep}{5.1mm}
   \begin{tabular}{@{}cc@{}}
        \hline
        Pseudo-observable                               & Value ($68\%$ gaussian)        \\
        \hline
        $|r_\perp^{J/\psi}|$                            & $(2.027 \pm 0.190) \cdot 10^{-3}$  \\
        $|r_\|^{J/\psi}|$                               & $(1.713 \pm 0.260) \cdot 10^{-3}$   \\
        $|r_0^{J/\psi}|$                                & $(2.303 \pm 0.357) \cdot 10^{-3}$   \\
        $\arg\{r_\perp^{J/\psi} r_{0}^{J/\psi*}\}$      & $+2.926 \pm 0.032$   \\
        $\arg\{r_\|^{J/\psi} r_{0}^{J/\psi*}\}$         & $-2.944 \pm 0.036$   \\[2mm]
        \hline
        $|r_\perp^{\psi(2S)}|$                          & $(1.06 \pm 0.21) \cdot 10^{-3}$  \\
        $|r_\|^{\psi(2S)}|$                             & $(0.98 \pm 0.18) \cdot 10^{-3}$   \\
        $|r_0^{\psi(2S)}|$                              & $(1.40 \pm 0.36) \cdot 10^{-3}$   \\
        $\arg\{r_\perp^{\psi(2S)} r_{0}^{\psi(2S)*}\}$    & $+2.799 \pm 0.314$   \\
        $\arg\{r_\|^{\psi(2S)} r_{0}^{\psi(2S)*}\}$       & $-2.815 \pm 0.403$   \\
        \hline
   \end{tabular}
    \caption{%
     Pseudo-observables from $B\to K^* \psi_n$.
    }
    \label{expPO}
\end{table}

The experimental constraints in \Sec{experimentalconstraints} are based on the
experimental pseudo-observables in \Eq{pseudoexp}, which are obtained by fit to
$B\to K^*\psi_n$ data.  For $B\to K^* J/\psi$ this data includes the
branching ratio as measured by Belle \cite{Chilikin:2014bkk}, as well as the
full set of angular observables $F_\perp$, $F_\parallel$, $\delta_\perp$ and
$\delta_\parallel$ measured by BaBar \cite{Aubert:2007hz} and LHCb
\cite{Aaij:2013cma}.  For $B\to K^* \psi(2S)$ the data includes the branching
ratio and the longitudinal polarization measured by Belle
\cite{Chilikin:2013tch}, and the full set of angular observables from BaBar
\cite{Aubert:2007hz}. For all measurements, correlations have been taken into
account where available. More recent results for the full angular
distributions, stemming from amplitude analyses that take into account
tetra-quark contributions \cite{Chilikin:2013tch,Chilikin:2014bkk}, are not used here.
The ansatz involving tetra-quark amplitudes is incompatible with the basis
of our analysis.
Although we expect to be able to use these additional results in future studies, this requires further dedicated work.

The relevant input to this fit are the CKM parameters, listed in \Tab{CKM},
and the form factors (see \Eq{r_lambda}). Since the experimental inputs
are sensitive to the form factors at $q^2 \leq 10\,\GeV^2$, we use
the combined fit to $K^*$-meson LCSR and Lattice results performed in~\Ref{Straub:2015ica}.
However, we \emph{double} the uncertainties quoted in~\cite{Straub:2015ica}
to ensure full agreement among the LCSRs and Lattice results.
Note that we do not account for correlations among the $r^{\psi_n}_\lambda$ due
to correlations among the form factor parameters.
The results for the pseudo-observables are given in \Tab{expPO}. The two sets of
observables for $J/\psi$ and $\psi(2S)$ are correlated with correlation
matrices:
{
\small
\eqa{
\rho_{J/\psi} &=& \left(
        \begin{array}{rrrrr}
             1.000 &  0.786 &  0.213 & -0.007 & -0.026 \\
                   &  1.000 &  0.177 &  0.003 &  0.011 \\
                   &        &  1.000 & -0.003 & -0.004 \\
                   &        &        &  1.000 &  0.652 \\
                   &        &        &        &  1.000 \\
        \end{array}
    \right)\,,\\[3mm]
\rho_{\psi(2S)} &=& \left(
        \begin{array}{rrrrr}
             1.000 & -0.116 &  0.233 & -0.222 & -0.204 \\
                   &  1.000 &  0.252 &  0.204 &  0.173 \\
                   &        &  1.000 & -0.007 &  0.008 \\
                   &        &        &  1.000 &  0.679 \\
                   &        &        &        &  1.000 \\
        \end{array}
    \right)\,.\qquad
}
}
In both cases, the mean and standard deviations have been obtained from a fit to
$10^6$ samples of the posterior predictive distributions. On the other hand, the
correlation coefficients have been obtained from the sample covariance of these $10^6$
samples. It is noteworthy that none of
the coefficients exceeds a level of $78\%$ for the $J/\psi$ and $68\%$ for
the $\psi(2S)$, respectively.

The theory constraints in \Sec{theoryconstraints} are based on pseudo-observables at four different points
at spacelike~$q^2$. The derived values including uncertainties and correlations are listed in \Tab{theoryPO}.

\begin{table*}
    \renewcommand{\arraystretch}{1.5}
    \begin{tabular}{cc|rrrr|rrrr|rrrr}
        \hline
        {}
            &
            & \multicolumn{4}{c|}{${\rm Re}[{\H}_\perp]/\F_\perp$}
            & \multicolumn{4}{c|}{${\rm Re}[{\H}_\|]/\F_\|$}
            & \multicolumn{4}{c }{${\rm Re}[{\H}_0]/\F_0$} \\
        {}
            & $q^2$
            & -7.0 & -5.0 & -3.0 & -1.0
            & -7.0 & -5.0 & -3.0 & -1.0
            & -7.0 & -5.0 & -3.0 & -1.0 \\
        \hline
        {}
            & $\mu$
            &  6.656 & 4.878 & 4.076 & 3.750
            &  6.033 & 4.384 & 3.728 & 3.586
            & -1.997 & 1.596 & 1.818 & 0.768 \\
        \hline
        {}
            & $\sigma$
            & 2.553 & 1.048 & 0.621 & 0.561
            & 2.446 & 0.971 & 0.575 & 0.538
            & 4.077 & 1.368 & 0.472 & 0.125 \\
        \hline
        \hline
        {}
            &
            & \multicolumn{4}{c|}{${\rm Im}[{\H}_\perp]/\F_\perp$}
            & \multicolumn{4}{c|}{${\rm Im}[{\H}_\|]/\F_\|$}
            & \multicolumn{4}{c }{${\rm Im}[{\H}_0]/\F_0$} \\
        {}
            & $q^2$
            & -7.0 & -5.0 & -3.0 & -1.0
            & -7.0 & -5.0 & -3.0 & -1.0
            & -7.0 & -5.0 & -3.0 & -1.0 \\
        \hline
        {}
            & $\mu$
            &  1.581 & 1.294 & 1.291 & 1.380
            &  1.517 & 1.246 & 1.257 & 1.366
            &  6.328 & 1.970 & 0.583 & 0.136 \\
        \hline
        {}
            & $\sigma$
            & 0.835 & 0.610 & 0.565 & 0.585
            & 0.803 & 0.590 & 0.553 & 0.581
            & 7.411 & 2.107 & 0.528 & 0.082 \\
        \hline
    \end{tabular}
\caption{Mean values $\mu_i$ (in units of $10^{-4}$), and standard deviations $\sigma_i$ (in units of $10^{-4}$)
of the theory constraints at negative $q^2$ (in units of $\GeV^2$).}
\label{theoryPO}
\end{table*}
\begin{figure*}
    \subfigure[\label{ffV}]{\includegraphics[width=.49\textwidth]{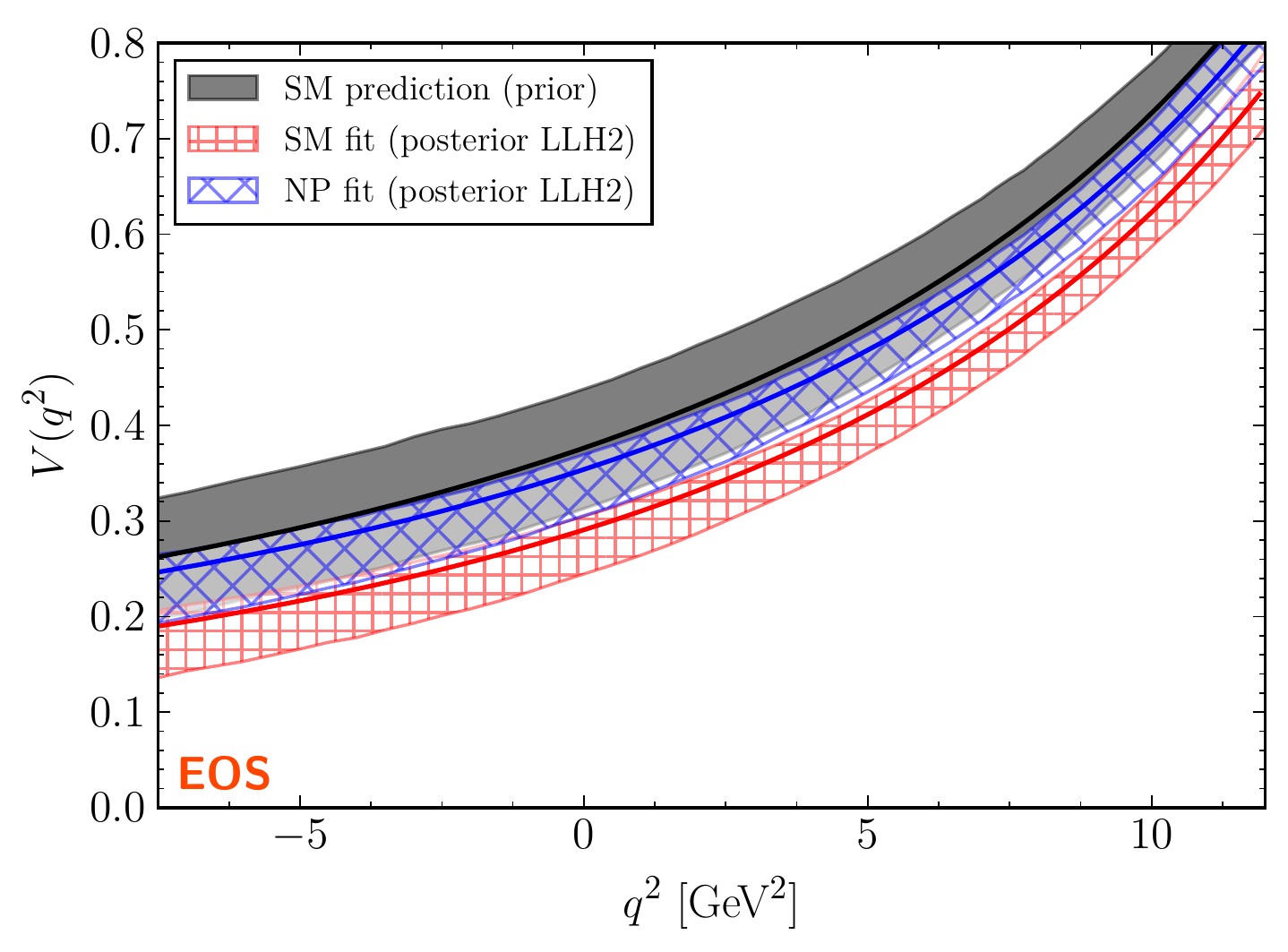}}
    \subfigure[\label{symrelA1V}]{\includegraphics[width=.49\textwidth]{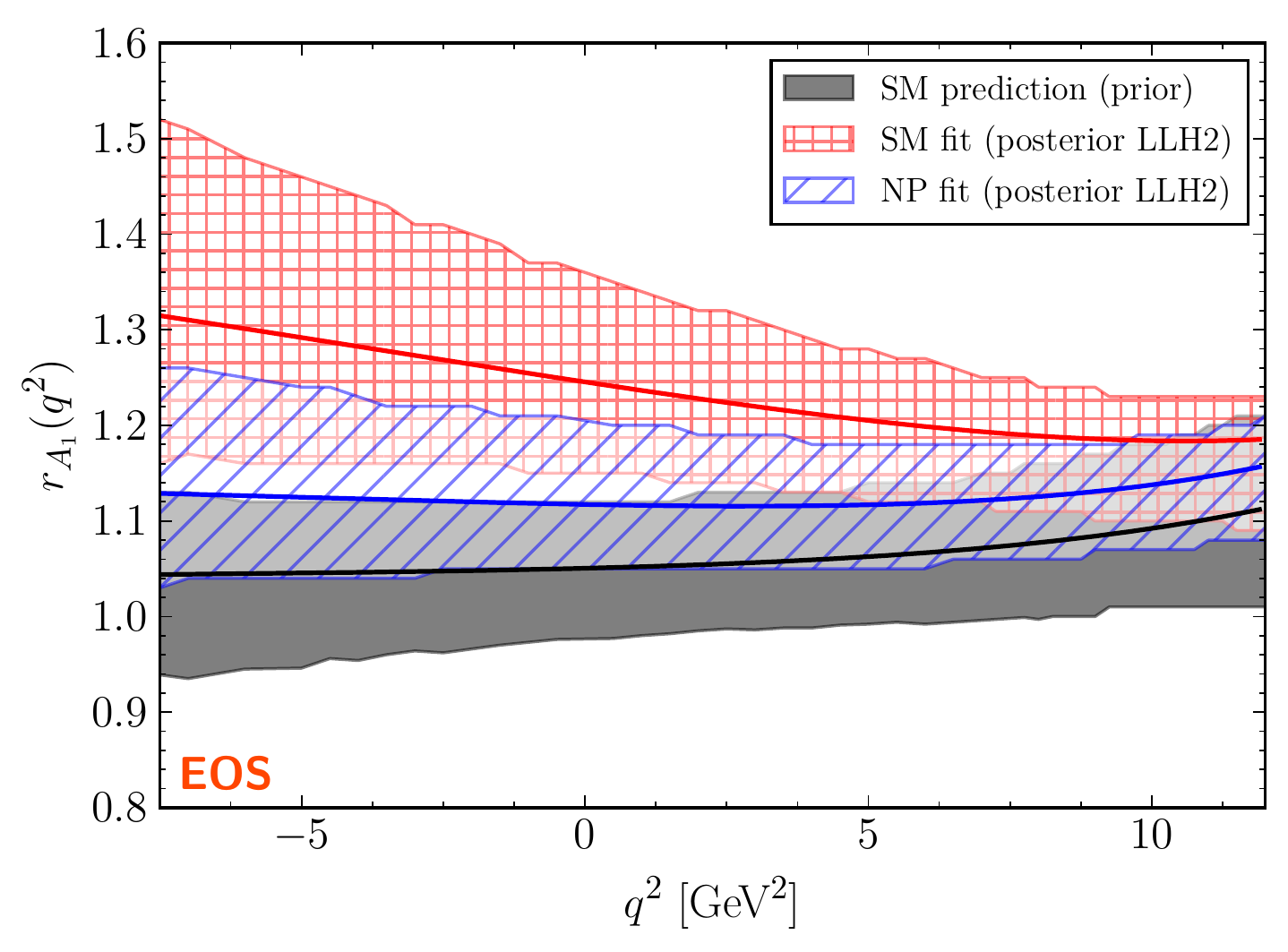}}
    \caption{A-priori predictions of the form factor $V$ and the symmetry ratio $r_{A_1}$ as functions of $q^2$,
    and comparison with two a-posteriori results.}
\end{figure*}

\begin{figure*}
    \begin{tabular}{cc}
        \subfigure[]{\includegraphics[width=.49\textwidth]{plots/fig-btokstarccbar-re-ratio-perp-paramA-z.pdf}} &
        \subfigure[]{\includegraphics[width=.49\textwidth]{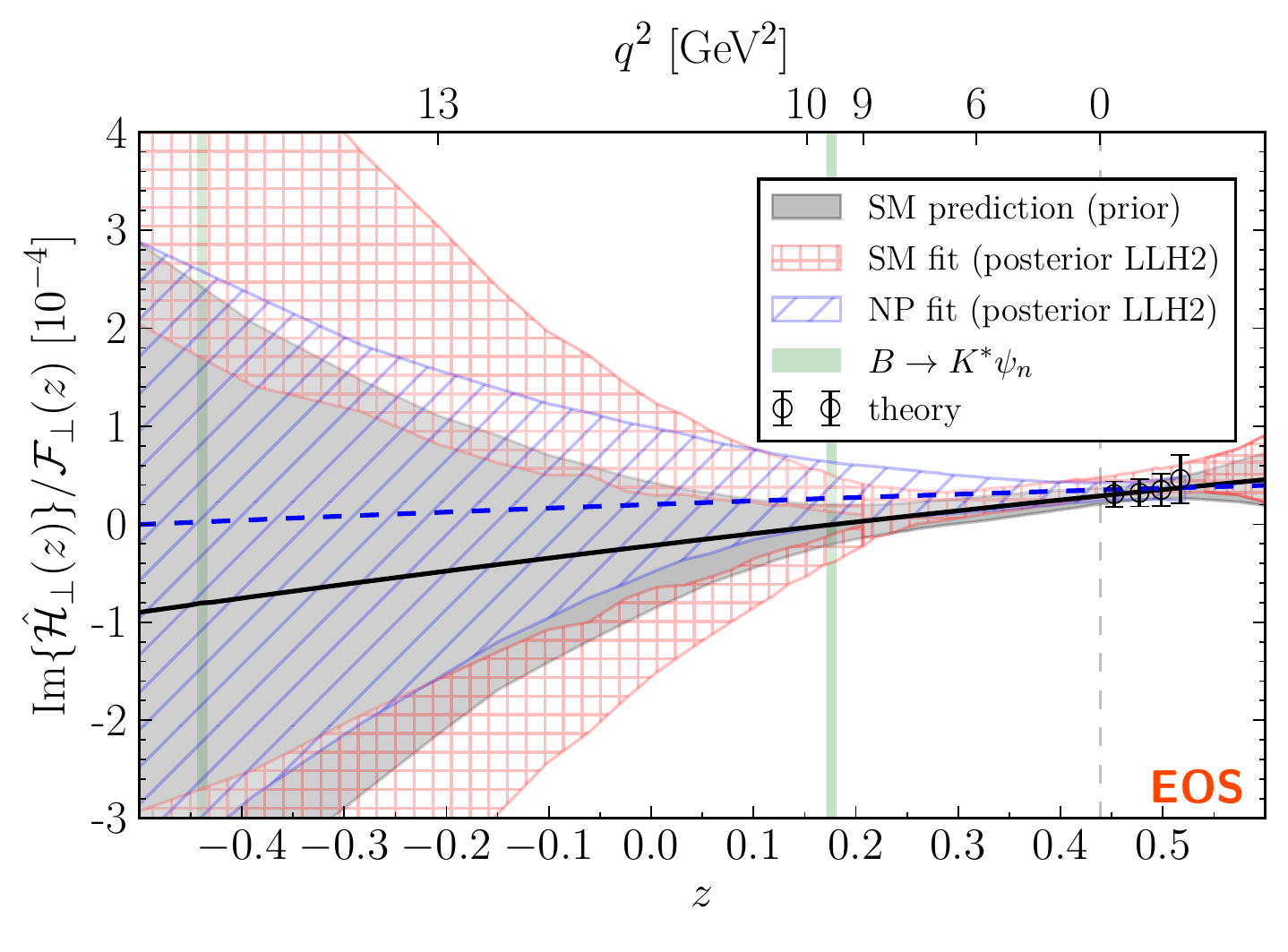}} \\
        \subfigure[]{\includegraphics[width=.49\textwidth]{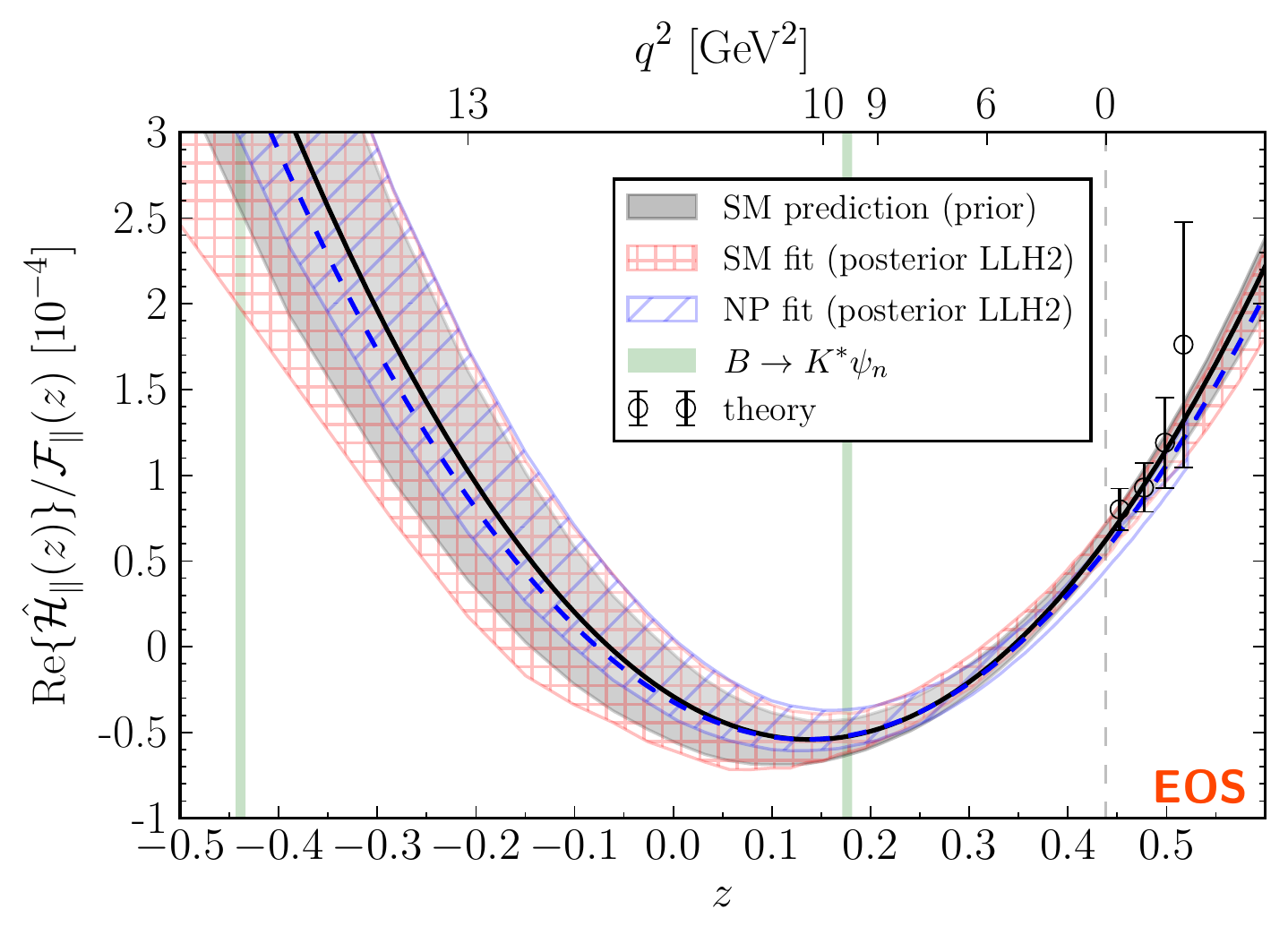}} &
        \subfigure[]{\includegraphics[width=.49\textwidth]{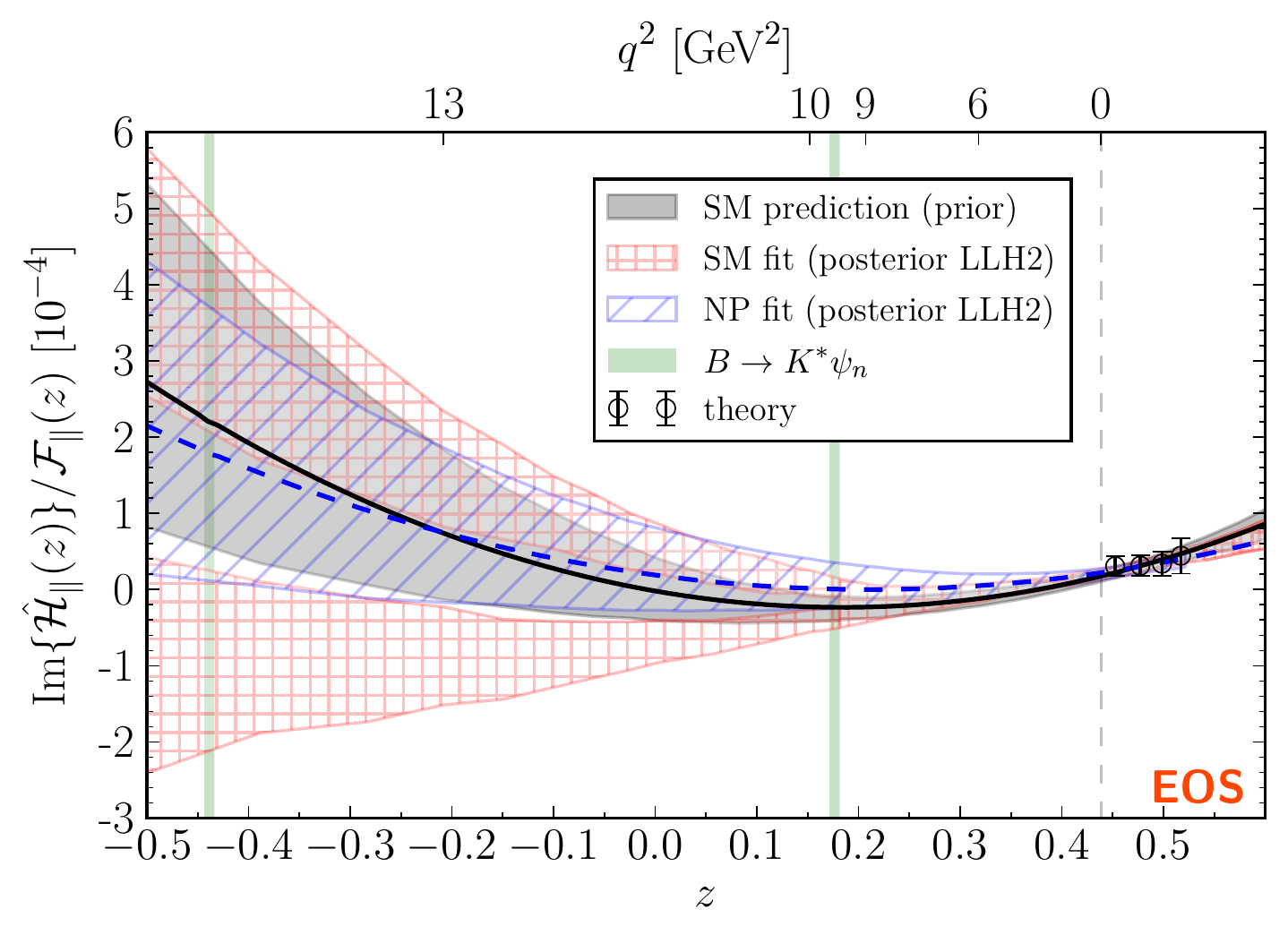}} \\
        \subfigure[]{\includegraphics[width=.49\textwidth]{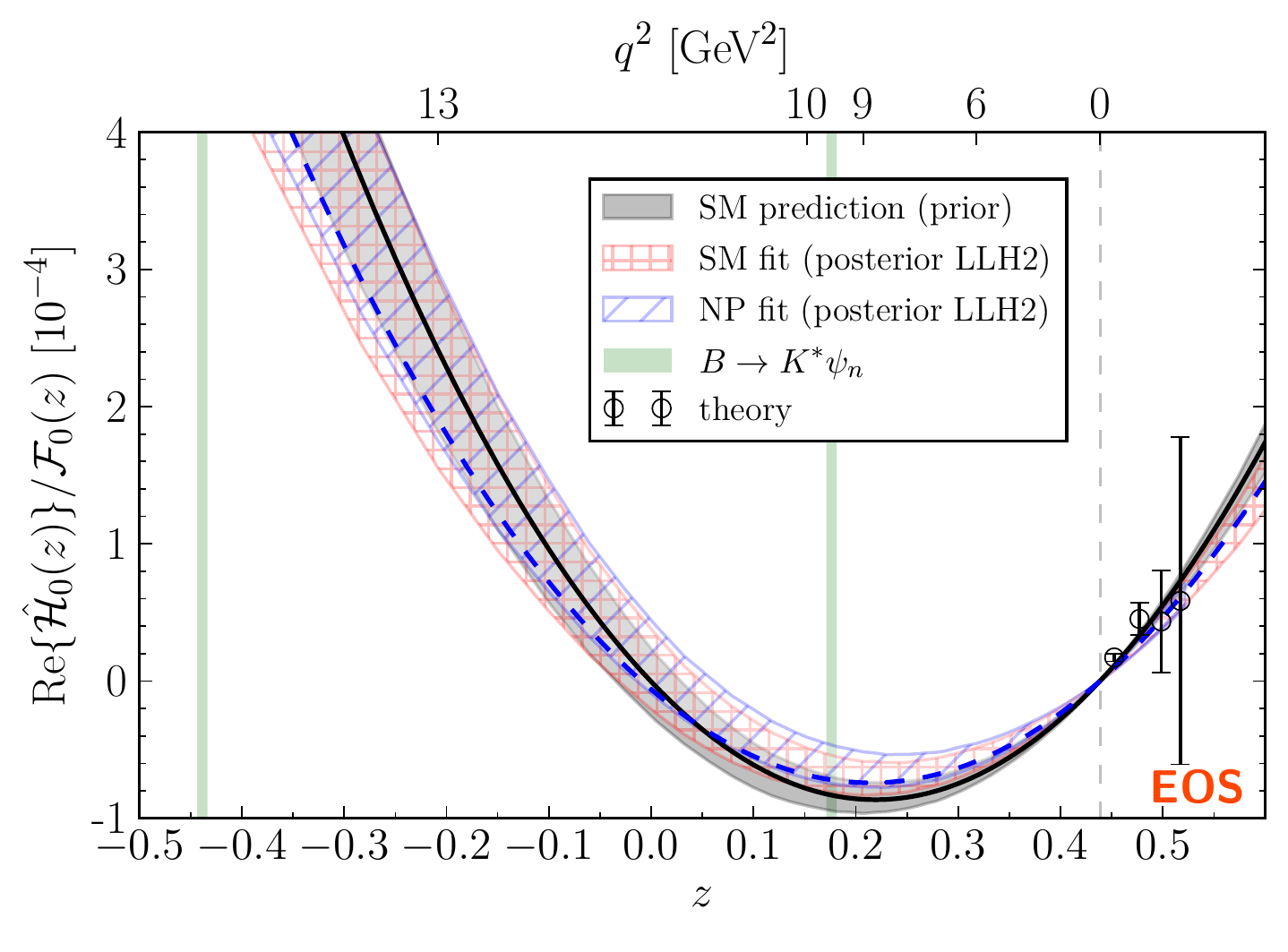}} &
        \subfigure[]{\includegraphics[width=.49\textwidth]{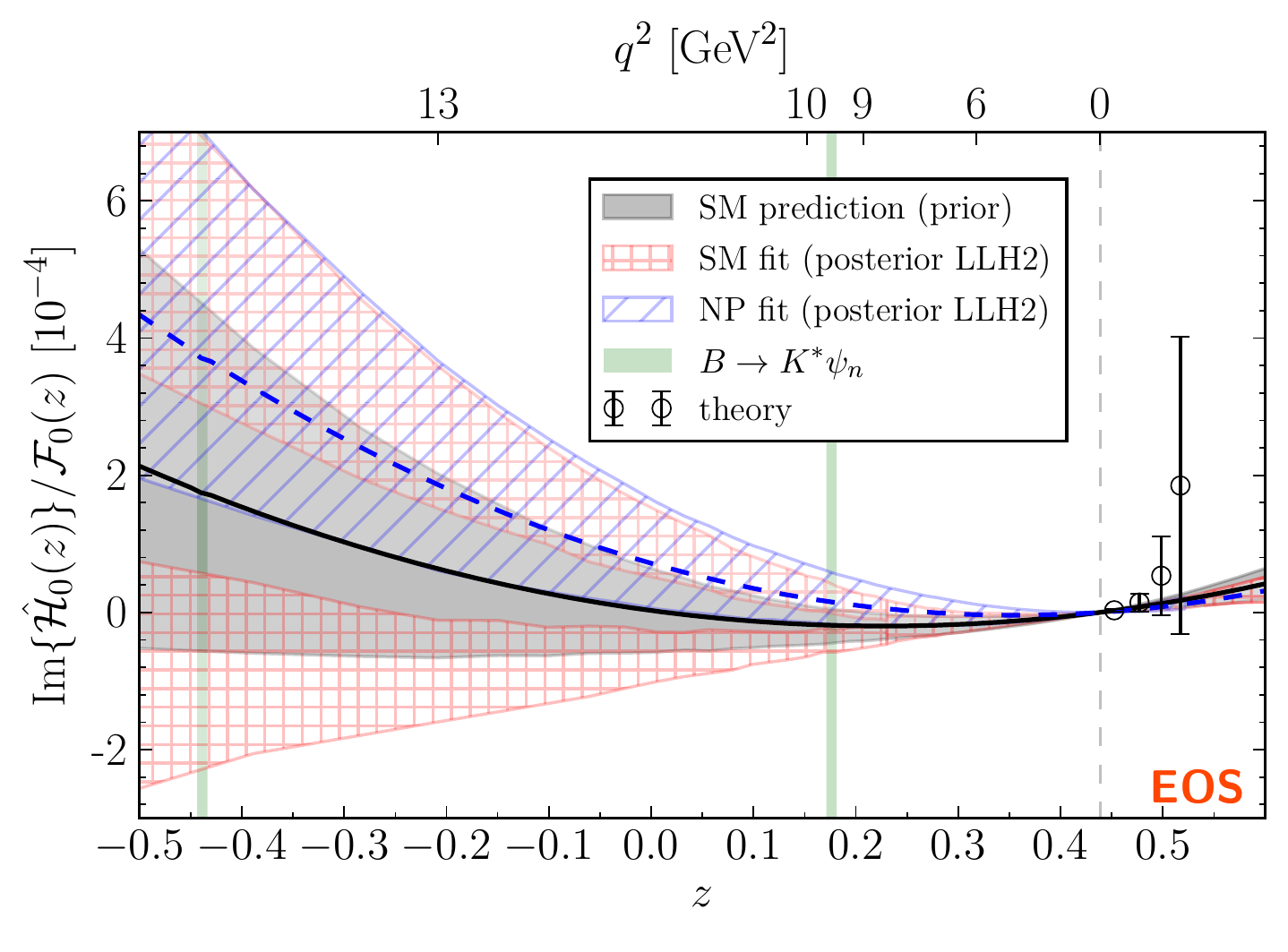}}
    \end{tabular}
    \caption{A-priori predictions of the ratios $\mathcal{H}_\lambda/\mathcal{F}_\lambda$ as functions of $z$,
    and comparison with two a-posteriori results. Note that the SM fit posterior is multi-modal, and we therefore
    choose not to illustrate its best-fit curves.}
\label{allplots}
\end{figure*}

Nominally, for $K=2$ the fit would involve $18$ real-valued parameters $\alpha_k^{(\lambda)}$.
Using the property that the longitudinal correlator must vanish at zero momentum transfer $q^2=0$,
$\hat{\H}_0(z(q^2 = 0)) = 0$, we can reduce the number of parameters by $2$ through the
replacement
\eq{
\begin{aligned}
    \alpha_0^{(0)} & \to \alpha_0^{\prime(0)} \equiv -z(0)\, \alpha_0^{(0)}\ , \\
    \alpha_1^{(0)} & \to \alpha_1^{\prime(0)} \equiv \alpha_0^{(0)} -z(0)\, \alpha_1^{(0)}\ , \\
    \alpha_2^{(0)} & \to \alpha_2^{\prime(0)} \equiv \alpha_1^{(0)}\ .
\end{aligned}
}
From the fit of the parameters $\alpha_k^{(\lambda)}$ to the theory constraints and the
pseudo-observables $r_\lambda^{\psi_n}$ we find all 1D-marginalized posteriors
to be reasonably symmetric around their modes. As for the pseudo-observables,
we obtain means and standard deviations from fits to $10^6$ samples of the PDF,
while the correlation coefficients are obtained through computation of the
sample covariance. Our results are summarized in \Tab{alphak}.  Finally, the
set of predictive distributions for the ratios $\H_\lambda/\F_\lambda$ are
shown in \Fig{allplots} for all transversities, including real and imaginary parts.

A key result of our analysis is that modifications to the correlator parameters cannot bring the
theory predictions and measurements into better agreement. However, modifications to the form factor parameters
can reduce the tensions, albeit not completely. We show in \Fig{ffV} the impact on the form factor $V(q^2)$,
which is the one most affected. Moreover, since the form factor $A_1$ is almost unaffected, we find
that the SM fit prefers a substantial violation of the large-recoil symmetry relation
\cite{Charles:1998dr,Beneke:2000wa} involving $V$ and $A_1$,
\eq{
    r_{A_1}(q^2) \equiv \frac{(M_B + M_{K^*})^2}{2 M_B E_{K^*}(q^2)}\, \frac{A_1(q^2)}{V(q^2)}\,,
}
as shown in \Fig{symrelA1V}.

\bigskip

\bibliography{references.bib}

\begin{thebibliography}{51}%
\makeatletter
\providecommand \@ifxundefined [1]{%
 \@ifx{#1\undefined}
}%
\providecommand \@ifnum [1]{%
 \ifnum #1\expandafter \@firstoftwo
 \else \expandafter \@secondoftwo
 \fi
}%
\providecommand \@ifx [1]{%
 \ifx #1\expandafter \@firstoftwo
 \else \expandafter \@secondoftwo
 \fi
}%
\providecommand \natexlab [1]{#1}%
\providecommand \enquote  [1]{``#1''}%
\providecommand \bibnamefont  [1]{#1}%
\providecommand \bibfnamefont [1]{#1}%
\providecommand \citenamefont [1]{#1}%
\providecommand \href@noop [0]{\@secondoftwo}%
\providecommand \href [0]{\begingroup \@sanitize@url \@href}%
\providecommand \@href[1]{\@@startlink{#1}\@@href}%
\providecommand \@@href[1]{\endgroup#1\@@endlink}%
\providecommand \@sanitize@url [0]{\catcode `\\12\catcode `\$12\catcode
  `\&12\catcode `\#12\catcode `\^12\catcode `\_12\catcode `\%12\relax}%
\providecommand \@@startlink[1]{}%
\providecommand \@@endlink[0]{}%
\providecommand \url  [0]{\begingroup\@sanitize@url \@url }%
\providecommand \@url [1]{\endgroup\@href {#1}{\urlprefix }}%
\providecommand \urlprefix  [0]{URL }%
\providecommand \Eprint [0]{\href }%
\providecommand \doibase [0]{http://dx.doi.org/}%
\providecommand \selectlanguage [0]{\@gobble}%
\providecommand \bibinfo  [0]{\@secondoftwo}%
\providecommand \bibfield  [0]{\@secondoftwo}%
\providecommand \translation [1]{[#1]}%
\providecommand \BibitemOpen [0]{}%
\providecommand \bibitemStop [0]{}%
\providecommand \bibitemNoStop [0]{.\EOS\space}%
\providecommand \EOS [0]{\spacefactor3000\relax}%
\providecommand \BibitemShut  [1]{\csname bibitem#1\endcsname}%
\let\auto@bib@innerbib\@empty
\bibitem [{\citenamefont {Ball}\ and\ \citenamefont
  {Braun}(1998)}]{Ball:1998kk}%
  \BibitemOpen
  \bibfield  {author} {\bibinfo {author} {\bibfnamefont {Patricia}\
  \bibnamefont {Ball}}\ and\ \bibinfo {author} {\bibfnamefont {Vladimir~M.}\
  \bibnamefont {Braun}},\ }\bibfield  {title} {\enquote {\bibinfo {title}
  {{Exclusive semileptonic and rare B meson decays in QCD}},}\ }\href {\doibase
  10.1103/PhysRevD.58.094016} {\bibfield  {journal} {\bibinfo  {journal} {Phys.
  Rev.}\ }\textbf {\bibinfo {volume} {D58}},\ \bibinfo {pages} {094016}
  (\bibinfo {year} {1998})},\ \Eprint {http://arxiv.org/abs/hep-ph/9805422}
  {arXiv:hep-ph/9805422 [hep-ph]} \BibitemShut {NoStop}%
\bibitem [{\citenamefont {Khodjamirian}\ \emph {et~al.}(2007)\citenamefont
  {Khodjamirian}, \citenamefont {Mannel},\ and\ \citenamefont
  {Offen}}]{Khodjamirian:2006st}%
  \BibitemOpen
  \bibfield  {author} {\bibinfo {author} {\bibfnamefont {Alexander}\
  \bibnamefont {Khodjamirian}}, \bibinfo {author} {\bibfnamefont {Thomas}\
  \bibnamefont {Mannel}}, \ and\ \bibinfo {author} {\bibfnamefont {Nils}\
  \bibnamefont {Offen}},\ }\bibfield  {title} {\enquote {\bibinfo {title}
  {{Form-factors from light-cone sum rules with B-meson distribution
  amplitudes}},}\ }\href {\doibase 10.1103/PhysRevD.75.054013} {\bibfield
  {journal} {\bibinfo  {journal} {Phys. Rev.}\ }\textbf {\bibinfo {volume}
  {D75}},\ \bibinfo {pages} {054013} (\bibinfo {year} {2007})},\ \Eprint
  {http://arxiv.org/abs/hep-ph/0611193} {arXiv:hep-ph/0611193 [hep-ph]}
  \BibitemShut {NoStop}%
\bibitem [{\citenamefont {Becirevic}\ \emph {et~al.}(2007)\citenamefont
  {Becirevic}, \citenamefont {Lubicz},\ and\ \citenamefont
  {Mescia}}]{Becirevic:2006nm}%
  \BibitemOpen
  \bibfield  {author} {\bibinfo {author} {\bibfnamefont {Damir}\ \bibnamefont
  {Becirevic}}, \bibinfo {author} {\bibfnamefont {Vittorio}\ \bibnamefont
  {Lubicz}}, \ and\ \bibinfo {author} {\bibfnamefont {Federico}\ \bibnamefont
  {Mescia}},\ }\bibfield  {title} {\enquote {\bibinfo {title} {{An Estimate of
  the $B \to K^* \gamma$ form factor}},}\ }\href {\doibase
  10.1016/j.nuclphysb.2007.01.032} {\bibfield  {journal} {\bibinfo  {journal}
  {Nucl. Phys.}\ }\textbf {\bibinfo {volume} {B769}},\ \bibinfo {pages}
  {31--43} (\bibinfo {year} {2007})},\ \Eprint
  {http://arxiv.org/abs/hep-ph/0611295} {arXiv:hep-ph/0611295 [hep-ph]}
  \BibitemShut {NoStop}%
\bibitem [{\citenamefont {Horgan}\ \emph {et~al.}(2014)\citenamefont {Horgan},
  \citenamefont {Liu}, \citenamefont {Meinel},\ and\ \citenamefont
  {Wingate}}]{Horgan:2013hoa}%
  \BibitemOpen
  \bibfield  {author} {\bibinfo {author} {\bibfnamefont {Ronald~R.}\
  \bibnamefont {Horgan}}, \bibinfo {author} {\bibfnamefont {Zhaofeng}\
  \bibnamefont {Liu}}, \bibinfo {author} {\bibfnamefont {Stefan}\ \bibnamefont
  {Meinel}}, \ and\ \bibinfo {author} {\bibfnamefont {Matthew}\ \bibnamefont
  {Wingate}},\ }\bibfield  {title} {\enquote {\bibinfo {title} {{Lattice QCD
  calculation of form factors describing the rare decays $B \to K^* \ell^+
  \ell^-$ and $B_s \to \phi \ell^+ \ell^-$}},}\ }\href {\doibase
  10.1103/PhysRevD.89.094501} {\bibfield  {journal} {\bibinfo  {journal} {Phys.
  Rev.}\ }\textbf {\bibinfo {volume} {D89}},\ \bibinfo {pages} {094501}
  (\bibinfo {year} {2014})},\ \Eprint {http://arxiv.org/abs/1310.3722}
  {arXiv:1310.3722 [hep-lat]} \BibitemShut {NoStop}%
\bibitem [{\citenamefont {Descotes-Genon}\ \emph
  {et~al.}(2013{\natexlab{a}})\citenamefont {Descotes-Genon}, \citenamefont
  {Hurth}, \citenamefont {Matias},\ and\ \citenamefont
  {Virto}}]{Descotes-Genon:2013vna}%
  \BibitemOpen
  \bibfield  {author} {\bibinfo {author} {\bibfnamefont {Sebastien}\
  \bibnamefont {Descotes-Genon}}, \bibinfo {author} {\bibfnamefont {Tobias}\
  \bibnamefont {Hurth}}, \bibinfo {author} {\bibfnamefont {Joaquim}\
  \bibnamefont {Matias}}, \ and\ \bibinfo {author} {\bibfnamefont {Javier}\
  \bibnamefont {Virto}},\ }\bibfield  {title} {\enquote {\bibinfo {title}
  {{Optimizing the basis of $B\to K^*ll$ observables in the full kinematic
  range}},}\ }\href {\doibase 10.1007/JHEP05(2013)137} {\bibfield  {journal}
  {\bibinfo  {journal} {JHEP}\ }\textbf {\bibinfo {volume} {05}},\ \bibinfo
  {pages} {137} (\bibinfo {year} {2013}{\natexlab{a}})},\ \Eprint
  {http://arxiv.org/abs/1303.5794} {arXiv:1303.5794 [hep-ph]} \BibitemShut
  {NoStop}%
\bibitem [{\citenamefont {Bharucha}\ \emph {et~al.}(2016)\citenamefont
  {Bharucha}, \citenamefont {Straub},\ and\ \citenamefont
  {Zwicky}}]{Straub:2015ica}%
  \BibitemOpen
  \bibfield  {author} {\bibinfo {author} {\bibfnamefont {Aoife}\ \bibnamefont
  {Bharucha}}, \bibinfo {author} {\bibfnamefont {David~M.}\ \bibnamefont
  {Straub}}, \ and\ \bibinfo {author} {\bibfnamefont {Roman}\ \bibnamefont
  {Zwicky}},\ }\bibfield  {title} {\enquote {\bibinfo {title} {{$B\to
  V\ell^+\ell^-$ in the Standard Model from light-cone sum rules}},}\ }\href
  {\doibase 10.1007/JHEP08(2016)098} {\bibfield  {journal} {\bibinfo  {journal}
  {JHEP}\ }\textbf {\bibinfo {volume} {08}},\ \bibinfo {pages} {098} (\bibinfo
  {year} {2016})},\ \Eprint {http://arxiv.org/abs/1503.05534} {arXiv:1503.05534
  [hep-ph]} \BibitemShut {NoStop}%
\bibitem [{\citenamefont {Hansen}\ and\ \citenamefont
  {Sharpe}(2012)}]{Hansen:2012tf}%
  \BibitemOpen
  \bibfield  {author} {\bibinfo {author} {\bibfnamefont {Maxwell~T.}\
  \bibnamefont {Hansen}}\ and\ \bibinfo {author} {\bibfnamefont {Stephen~R.}\
  \bibnamefont {Sharpe}},\ }\bibfield  {title} {\enquote {\bibinfo {title}
  {{Multiple-channel generalization of Lellouch-Luscher formula}},}\ }\href
  {\doibase 10.1103/PhysRevD.86.016007} {\bibfield  {journal} {\bibinfo
  {journal} {Phys. Rev.}\ }\textbf {\bibinfo {volume} {D86}},\ \bibinfo {pages}
  {016007} (\bibinfo {year} {2012})},\ \Eprint {http://arxiv.org/abs/1204.0826}
  {arXiv:1204.0826 [hep-lat]} \BibitemShut {NoStop}%
\bibitem [{\citenamefont {Briceño}\ \emph {et~al.}(2015)\citenamefont
  {Briceño}, \citenamefont {Hansen},\ and\ \citenamefont
  {Walker-Loud}}]{Briceno:2014uqa}%
  \BibitemOpen
  \bibfield  {author} {\bibinfo {author} {\bibfnamefont {Raúl~A.}\
  \bibnamefont {Briceño}}, \bibinfo {author} {\bibfnamefont {Maxwell~T.}\
  \bibnamefont {Hansen}}, \ and\ \bibinfo {author} {\bibfnamefont {André}\
  \bibnamefont {Walker-Loud}},\ }\bibfield  {title} {\enquote {\bibinfo {title}
  {{Multichannel 1 $\rightarrow$ 2 transition amplitudes in a finite
  volume}},}\ }\href {\doibase 10.1103/PhysRevD.91.034501} {\bibfield
  {journal} {\bibinfo  {journal} {Phys. Rev.}\ }\textbf {\bibinfo {volume}
  {D91}},\ \bibinfo {pages} {034501} (\bibinfo {year} {2015})},\ \Eprint
  {http://arxiv.org/abs/1406.5965} {arXiv:1406.5965 [hep-lat]} \BibitemShut
  {NoStop}%
\bibitem [{\citenamefont {Cheng}\ \emph {et~al.}(2017)\citenamefont {Cheng},
  \citenamefont {Khodjamirian},\ and\ \citenamefont {Virto}}]{Cheng:2017smj}%
  \BibitemOpen
  \bibfield  {author} {\bibinfo {author} {\bibfnamefont {Shan}\ \bibnamefont
  {Cheng}}, \bibinfo {author} {\bibfnamefont {Alexander}\ \bibnamefont
  {Khodjamirian}}, \ and\ \bibinfo {author} {\bibfnamefont {Javier}\
  \bibnamefont {Virto}},\ }\bibfield  {title} {\enquote {\bibinfo {title}
  {{$B\to\pi\pi$ Form Factors from Light-Cone Sum Rules with $B$-meson
  Distribution Amplitudes}},}\ }\href {\doibase 10.1007/JHEP05(2017)157}
  {\bibfield  {journal} {\bibinfo  {journal} {JHEP}\ }\textbf {\bibinfo
  {volume} {05}},\ \bibinfo {pages} {157} (\bibinfo {year} {2017})},\ \Eprint
  {http://arxiv.org/abs/1701.01633} {arXiv:1701.01633 [hep-ph]} \BibitemShut
  {NoStop}%
\bibitem [{\citenamefont {B{\"o}er}\ \emph {et~al.}(2017)\citenamefont
  {B{\"o}er}, \citenamefont {Feldmann},\ and\ \citenamefont {van
  Dyk}}]{Boer:2016iez}%
  \BibitemOpen
  \bibfield  {author} {\bibinfo {author} {\bibfnamefont {Philipp}\ \bibnamefont
  {B{\"o}er}}, \bibinfo {author} {\bibfnamefont {Thorsten}\ \bibnamefont
  {Feldmann}}, \ and\ \bibinfo {author} {\bibfnamefont {Danny}\ \bibnamefont
  {van Dyk}},\ }\bibfield  {title} {\enquote {\bibinfo {title} {{QCD
  Factorization for $B \to \pi\pi\ell\nu$ Decays at Large Dipion Masses}},}\
  }\href {\doibase 10.1007/JHEP02(2017)133} {\bibfield  {journal} {\bibinfo
  {journal} {JHEP}\ }\textbf {\bibinfo {volume} {02}},\ \bibinfo {pages} {133}
  (\bibinfo {year} {2017})},\ \Eprint {http://arxiv.org/abs/1608.07127}
  {arXiv:1608.07127 [hep-ph]} \BibitemShut {NoStop}%
\bibitem [{\citenamefont {Wang}\ and\ \citenamefont
  {Shen}(2015)}]{Wang:2015vgv}%
  \BibitemOpen
  \bibfield  {author} {\bibinfo {author} {\bibfnamefont {Yu-Ming}\ \bibnamefont
  {Wang}}\ and\ \bibinfo {author} {\bibfnamefont {Yue-Long}\ \bibnamefont
  {Shen}},\ }\bibfield  {title} {\enquote {\bibinfo {title} {{QCD corrections
  to $B\to \pi$ form factors from light-cone sum rules}},}\ }\href {\doibase
  10.1016/j.nuclphysb.2015.07.016} {\bibfield  {journal} {\bibinfo  {journal}
  {Nucl. Phys.}\ }\textbf {\bibinfo {volume} {B898}},\ \bibinfo {pages}
  {563--604} (\bibinfo {year} {2015})},\ \Eprint
  {http://arxiv.org/abs/1506.00667} {arXiv:1506.00667 [hep-ph]} \BibitemShut
  {NoStop}%
\bibitem [{\citenamefont {Beneke}\ \emph {et~al.}(2001)\citenamefont {Beneke},
  \citenamefont {Feldmann},\ and\ \citenamefont {Seidel}}]{Beneke:2001at}%
  \BibitemOpen
  \bibfield  {author} {\bibinfo {author} {\bibfnamefont {M.}~\bibnamefont
  {Beneke}}, \bibinfo {author} {\bibfnamefont {T.}~\bibnamefont {Feldmann}}, \
  and\ \bibinfo {author} {\bibfnamefont {D.}~\bibnamefont {Seidel}},\
  }\bibfield  {title} {\enquote {\bibinfo {title} {{Systematic approach to
  exclusive $B \to V \ell^+ \ell^-$, $V \gamma$ decays}},}\ }\href {\doibase
  10.1016/S0550-3213(01)00366-2} {\bibfield  {journal} {\bibinfo  {journal}
  {Nucl. Phys.}\ }\textbf {\bibinfo {volume} {B612}},\ \bibinfo {pages}
  {25--58} (\bibinfo {year} {2001})},\ \Eprint
  {http://arxiv.org/abs/hep-ph/0106067} {arXiv:hep-ph/0106067 [hep-ph]}
  \BibitemShut {NoStop}%
\bibitem [{\citenamefont {Khodjamirian}\ \emph {et~al.}(2010)\citenamefont
  {Khodjamirian}, \citenamefont {Mannel}, \citenamefont {Pivovarov},\ and\
  \citenamefont {Wang}}]{Khodjamirian:2010vf}%
  \BibitemOpen
  \bibfield  {author} {\bibinfo {author} {\bibfnamefont {A.}~\bibnamefont
  {Khodjamirian}}, \bibinfo {author} {\bibfnamefont {Th.}\ \bibnamefont
  {Mannel}}, \bibinfo {author} {\bibfnamefont {A.~A.}\ \bibnamefont
  {Pivovarov}}, \ and\ \bibinfo {author} {\bibfnamefont {Y.~M.}\ \bibnamefont
  {Wang}},\ }\bibfield  {title} {\enquote {\bibinfo {title} {{Charm-loop effect
  in $B \to K^{(*)} \ell^{+} \ell^{-}$ and $B\to K^*\gamma$}},}\ }\href
  {\doibase 10.1007/JHEP09(2010)089} {\bibfield  {journal} {\bibinfo  {journal}
  {JHEP}\ }\textbf {\bibinfo {volume} {09}},\ \bibinfo {pages} {089} (\bibinfo
  {year} {2010})},\ \Eprint {http://arxiv.org/abs/1006.4945} {arXiv:1006.4945
  [hep-ph]} \BibitemShut {NoStop}%
\bibitem [{\citenamefont {Asatryan}\ \emph {et~al.}(2002)\citenamefont
  {Asatryan}, \citenamefont {Asatrian}, \citenamefont {Greub},\ and\
  \citenamefont {Walker}}]{Asatryan:2001zw}%
  \BibitemOpen
  \bibfield  {author} {\bibinfo {author} {\bibfnamefont {H.~H.}\ \bibnamefont
  {Asatryan}}, \bibinfo {author} {\bibfnamefont {H.~M.}\ \bibnamefont
  {Asatrian}}, \bibinfo {author} {\bibfnamefont {C.}~\bibnamefont {Greub}}, \
  and\ \bibinfo {author} {\bibfnamefont {M.}~\bibnamefont {Walker}},\
  }\bibfield  {title} {\enquote {\bibinfo {title} {{Calculation of two loop
  virtual corrections to $b \to s \ell^+ \ell^-$ in the standard model}},}\
  }\href {\doibase 10.1103/PhysRevD.65.074004} {\bibfield  {journal} {\bibinfo
  {journal} {Phys. Rev.}\ }\textbf {\bibinfo {volume} {D65}},\ \bibinfo {pages}
  {074004} (\bibinfo {year} {2002})},\ \Eprint
  {http://arxiv.org/abs/hep-ph/0109140} {arXiv:hep-ph/0109140 [hep-ph]}
  \BibitemShut {NoStop}%
\bibitem [{\citenamefont {Khodjamirian}\ \emph {et~al.}(2013)\citenamefont
  {Khodjamirian}, \citenamefont {Mannel},\ and\ \citenamefont
  {Wang}}]{Khodjamirian:2012rm}%
  \BibitemOpen
  \bibfield  {author} {\bibinfo {author} {\bibfnamefont {A.}~\bibnamefont
  {Khodjamirian}}, \bibinfo {author} {\bibfnamefont {Th.}\ \bibnamefont
  {Mannel}}, \ and\ \bibinfo {author} {\bibfnamefont {Y.~M.}\ \bibnamefont
  {Wang}},\ }\bibfield  {title} {\enquote {\bibinfo {title} {{$B \to K
  \ell^{+}\ell^{-}$ decay at large hadronic recoil}},}\ }\href {\doibase
  10.1007/JHEP02(2013)010} {\bibfield  {journal} {\bibinfo  {journal} {JHEP}\
  }\textbf {\bibinfo {volume} {02}},\ \bibinfo {pages} {010} (\bibinfo {year}
  {2013})},\ \Eprint {http://arxiv.org/abs/1211.0234} {arXiv:1211.0234
  [hep-ph]} \BibitemShut {NoStop}%
\bibitem [{\citenamefont {van Dyk}\ \emph
  {et~al.}(2017{\natexlab{a}})\citenamefont {van Dyk} \emph
  {et~al.}}]{EOS:web}%
  \BibitemOpen
  \bibfield  {author} {\bibinfo {author} {\bibfnamefont {D.}~\bibnamefont {van
  Dyk}} \emph {et~al.},\ }\href {https://eos.github.io/} {\enquote {\bibinfo
  {title} {{EOS -- A HEP Program for Flavour Observables}},}\ } (\bibinfo
  {year} {2017}{\natexlab{a}})\BibitemShut {NoStop}%
\bibitem [{\citenamefont {van Dyk}\ \emph
  {et~al.}(2017{\natexlab{b}})\citenamefont {van Dyk} \emph
  {et~al.}}]{EOS:release}%
  \BibitemOpen
  \bibfield  {author} {\bibinfo {author} {\bibfnamefont {D.}~\bibnamefont {van
  Dyk}} \emph {et~al.},\ }\href {\doibase 10.5281/zenodo.833765} {\enquote
  {\bibinfo {title} {{EOS (``analytic-btokstarll'' release)}},}\ } (\bibinfo
  {year} {2017}{\natexlab{b}})\BibitemShut {NoStop}%
\bibitem [{\citenamefont {Eden}\ \emph {et~al.}(1966)\citenamefont {Eden},
  \citenamefont {Landshoff}, \citenamefont {Olive},\ and\ \citenamefont
  {Polkinghorne}}]{Eden:1966dnq}%
  \BibitemOpen
  \bibfield  {author} {\bibinfo {author} {\bibfnamefont {Richard~John}\
  \bibnamefont {Eden}}, \bibinfo {author} {\bibfnamefont {Peter~V.}\
  \bibnamefont {Landshoff}}, \bibinfo {author} {\bibfnamefont {David~I.}\
  \bibnamefont {Olive}}, \ and\ \bibinfo {author} {\bibfnamefont
  {John~Charlton}\ \bibnamefont {Polkinghorne}},\ }\href@noop {} {\emph
  {\bibinfo {title} {{The analytic S-matrix}}}}\ (\bibinfo  {publisher}
  {Cambridge Univ. Press},\ \bibinfo {address} {Cambridge},\ \bibinfo {year}
  {1966})\BibitemShut {NoStop}%
\bibitem [{\citenamefont {Okubo}(1963)}]{Okubo:1963fa}%
  \BibitemOpen
  \bibfield  {author} {\bibinfo {author} {\bibfnamefont {S.}~\bibnamefont
  {Okubo}},\ }\bibfield  {title} {\enquote {\bibinfo {title} {{Phi meson and
  unitary symmetry model}},}\ }\href {\doibase 10.1016/S0375-9601(63)92548-9}
  {\bibfield  {journal} {\bibinfo  {journal} {Phys. Lett.}\ }\textbf {\bibinfo
  {volume} {5}},\ \bibinfo {pages} {165--168} (\bibinfo {year}
  {1963})}\BibitemShut {NoStop}%
\bibitem [{\citenamefont {Zweig}(1964)}]{Zweig:1964jf}%
  \BibitemOpen
  \bibfield  {author} {\bibinfo {author} {\bibfnamefont {G.}~\bibnamefont
  {Zweig}},\ }\bibfield  {title} {\enquote {\bibinfo {title} {{An $SU(3)$ model
  for strong interaction symmetry and its breaking. Version 2}},}\ }in\ \href
  {http://inspirehep.net/record/4674/files/cern-th-412.pdf} {\emph {\bibinfo
  {booktitle} {Developments in the quark theory of hadrons. Vol.\ 1.\ 1964 --
  1978}}},\ \bibinfo {editor} {edited by\ \bibinfo {editor} {\bibfnamefont
  {D.B.}\ \bibnamefont {Lichtenberg}}\ and\ \bibinfo {editor} {\bibfnamefont
  {Simon~Peter}\ \bibnamefont {Rosen}}}\ (\bibinfo {year} {1964})\ pp.\
  \bibinfo {pages} {22--101}\BibitemShut {NoStop}%
\bibitem [{\citenamefont {Iizuka}(1966)}]{Iizuka:1966fk}%
  \BibitemOpen
  \bibfield  {author} {\bibinfo {author} {\bibfnamefont {Jugoro}\ \bibnamefont
  {Iizuka}},\ }\bibfield  {title} {\enquote {\bibinfo {title} {{Systematics and
  phenomenology of meson family}},}\ }\href {\doibase 10.1143/PTPS.37.21}
  {\bibfield  {journal} {\bibinfo  {journal} {Prog. Theor. Phys. Suppl.}\
  }\textbf {\bibinfo {volume} {37}},\ \bibinfo {pages} {21--34} (\bibinfo
  {year} {1966})}\BibitemShut {NoStop}%
\bibitem [{\citenamefont {Boyd}\ \emph {et~al.}(1995)\citenamefont {Boyd},
  \citenamefont {Grinstein},\ and\ \citenamefont {Lebed}}]{Boyd:1995cf}%
  \BibitemOpen
  \bibfield  {author} {\bibinfo {author} {\bibfnamefont {C.~Glenn}\
  \bibnamefont {Boyd}}, \bibinfo {author} {\bibfnamefont {Benjamin}\
  \bibnamefont {Grinstein}}, \ and\ \bibinfo {author} {\bibfnamefont
  {Richard~F.}\ \bibnamefont {Lebed}},\ }\bibfield  {title} {\enquote {\bibinfo
  {title} {{Model independent extraction of $|V_{cb}|$ using dispersion
  relations}},}\ }\href {\doibase 10.1016/0370-2693(95)00480-9} {\bibfield
  {journal} {\bibinfo  {journal} {Phys. Lett.}\ }\textbf {\bibinfo {volume}
  {B353}},\ \bibinfo {pages} {306--312} (\bibinfo {year} {1995})},\ \Eprint
  {http://arxiv.org/abs/hep-ph/9504235} {arXiv:hep-ph/9504235 [hep-ph]}
  \BibitemShut {NoStop}%
\bibitem [{\citenamefont {Bourrely}\ \emph {et~al.}(2009)\citenamefont
  {Bourrely}, \citenamefont {Caprini},\ and\ \citenamefont
  {Lellouch}}]{Bourrely:2008za}%
  \BibitemOpen
  \bibfield  {author} {\bibinfo {author} {\bibfnamefont {Claude}\ \bibnamefont
  {Bourrely}}, \bibinfo {author} {\bibfnamefont {Irinel}\ \bibnamefont
  {Caprini}}, \ and\ \bibinfo {author} {\bibfnamefont {Laurent}\ \bibnamefont
  {Lellouch}},\ }\bibfield  {title} {\enquote {\bibinfo {title}
  {{Model-independent description of $B\to \pi\ell\nu$ decays and a
  determination of $|V_{ub}|$}},}\ }\href {\doibase 10.1103/PhysRevD.82.099902,
  10.1103/PhysRevD.79.013008} {\bibfield  {journal} {\bibinfo  {journal} {Phys.
  Rev.}\ }\textbf {\bibinfo {volume} {D79}},\ \bibinfo {pages} {013008}
  (\bibinfo {year} {2009})},\ \bibinfo {note} {[Erratum: Phys.
  Rev.D82,099902(2010)]},\ \Eprint {http://arxiv.org/abs/0807.2722}
  {arXiv:0807.2722 [hep-ph]} \BibitemShut {NoStop}%
\bibitem [{\citenamefont {Weinberg}(2005)}]{Weinberg:1995mt}%
  \BibitemOpen
  \bibfield  {author} {\bibinfo {author} {\bibfnamefont {Steven}\ \bibnamefont
  {Weinberg}},\ }\href@noop {} {\emph {\bibinfo {title} {{The Quantum theory of
  fields. Vol. 1: Foundations}}}}\ (\bibinfo  {publisher} {Cambridge University
  Press},\ \bibinfo {year} {2005})\BibitemShut {NoStop}%
\bibitem [{\citenamefont {Aubert}\ \emph {et~al.}(2005)\citenamefont {Aubert}
  \emph {et~al.}}]{Aubert:2004rz}%
  \BibitemOpen
  \bibfield  {author} {\bibinfo {author} {\bibfnamefont {Bernard}\ \bibnamefont
  {Aubert}} \emph {et~al.} (\bibinfo {collaboration} {BaBar}),\ }\bibfield
  {title} {\enquote {\bibinfo {title} {{Measurement of branching fractions and
  charge asymmetries for exclusive $B$ decays to charmonium}},}\ }\href
  {\doibase 10.1103/PhysRevLett.94.141801} {\bibfield  {journal} {\bibinfo
  {journal} {Phys. Rev. Lett.}\ }\textbf {\bibinfo {volume} {94}},\ \bibinfo
  {pages} {141801} (\bibinfo {year} {2005})},\ \Eprint
  {http://arxiv.org/abs/hep-ex/0412062} {arXiv:hep-ex/0412062 [hep-ex]}
  \BibitemShut {NoStop}%
\bibitem [{\citenamefont {Aubert}\ \emph {et~al.}(2007)\citenamefont {Aubert}
  \emph {et~al.}}]{Aubert:2007hz}%
  \BibitemOpen
  \bibfield  {author} {\bibinfo {author} {\bibfnamefont {Bernard}\ \bibnamefont
  {Aubert}} \emph {et~al.} (\bibinfo {collaboration} {BaBar}),\ }\bibfield
  {title} {\enquote {\bibinfo {title} {{Measurement of decay amplitudes of $B
  \to J/\psi K^{*}, \psi(2S) K^{*}$, and $\chi_{c1} K^{*}$ with an angular
  analysis}},}\ }\href {\doibase 10.1103/PhysRevD.76.031102} {\bibfield
  {journal} {\bibinfo  {journal} {Phys. Rev.}\ }\textbf {\bibinfo {volume}
  {D76}},\ \bibinfo {pages} {031102} (\bibinfo {year} {2007})},\ \Eprint
  {http://arxiv.org/abs/0704.0522} {arXiv:0704.0522 [hep-ex]} \BibitemShut
  {NoStop}%
\bibitem [{\citenamefont {Itoh}\ \emph {et~al.}(2005)\citenamefont {Itoh} \emph
  {et~al.}}]{Itoh:2005ks}%
  \BibitemOpen
  \bibfield  {author} {\bibinfo {author} {\bibfnamefont {R.}~\bibnamefont
  {Itoh}} \emph {et~al.} (\bibinfo {collaboration} {Belle}),\ }\bibfield
  {title} {\enquote {\bibinfo {title} {{Studies of CP violation in $B \to
  J/\psi K^*$ decays}},}\ }\href {\doibase 10.1103/PhysRevLett.95.091601}
  {\bibfield  {journal} {\bibinfo  {journal} {Phys. Rev. Lett.}\ }\textbf
  {\bibinfo {volume} {95}},\ \bibinfo {pages} {091601} (\bibinfo {year}
  {2005})},\ \Eprint {http://arxiv.org/abs/hep-ex/0504030}
  {arXiv:hep-ex/0504030 [hep-ex]} \BibitemShut {NoStop}%
\bibitem [{\citenamefont {Chilikin}\ \emph {et~al.}(2013)\citenamefont
  {Chilikin} \emph {et~al.}}]{Chilikin:2013tch}%
  \BibitemOpen
  \bibfield  {author} {\bibinfo {author} {\bibfnamefont {K.}~\bibnamefont
  {Chilikin}} \emph {et~al.} (\bibinfo {collaboration} {Belle}),\ }\bibfield
  {title} {\enquote {\bibinfo {title} {{Experimental constraints on the spin
  and parity of the $Z$(4430)$^+$}},}\ }\href {\doibase
  10.1103/PhysRevD.88.074026} {\bibfield  {journal} {\bibinfo  {journal} {Phys.
  Rev.}\ }\textbf {\bibinfo {volume} {D88}},\ \bibinfo {pages} {074026}
  (\bibinfo {year} {2013})},\ \Eprint {http://arxiv.org/abs/1306.4894}
  {arXiv:1306.4894 [hep-ex]} \BibitemShut {NoStop}%
\bibitem [{\citenamefont {Chilikin}\ \emph {et~al.}(2014)\citenamefont
  {Chilikin} \emph {et~al.}}]{Chilikin:2014bkk}%
  \BibitemOpen
  \bibfield  {author} {\bibinfo {author} {\bibfnamefont {K.}~\bibnamefont
  {Chilikin}} \emph {et~al.} (\bibinfo {collaboration} {Belle}),\ }\bibfield
  {title} {\enquote {\bibinfo {title} {{Observation of a new charged
  charmoniumlike state in $\bar{B}^0 \to J/\psi K^-\pi^+$ decays}},}\ }\href
  {\doibase 10.1103/PhysRevD.90.112009} {\bibfield  {journal} {\bibinfo
  {journal} {Phys. Rev.}\ }\textbf {\bibinfo {volume} {D90}},\ \bibinfo {pages}
  {112009} (\bibinfo {year} {2014})},\ \Eprint {http://arxiv.org/abs/1408.6457}
  {arXiv:1408.6457 [hep-ex]} \BibitemShut {NoStop}%
\bibitem [{\citenamefont {Aaij}\ \emph
  {et~al.}(2013{\natexlab{a}})\citenamefont {Aaij} \emph
  {et~al.}}]{Aaij:2013cma}%
  \BibitemOpen
  \bibfield  {author} {\bibinfo {author} {\bibfnamefont {R}~\bibnamefont
  {Aaij}} \emph {et~al.} (\bibinfo {collaboration} {LHCb}),\ }\bibfield
  {title} {\enquote {\bibinfo {title} {{Measurement of the polarization
  amplitudes in $B^0 \to J/\psi K^{*}(892)^0$ decays}},}\ }\href {\doibase
  10.1103/PhysRevD.88.052002} {\bibfield  {journal} {\bibinfo  {journal} {Phys.
  Rev.}\ }\textbf {\bibinfo {volume} {D88}},\ \bibinfo {pages} {052002}
  (\bibinfo {year} {2013}{\natexlab{a}})},\ \Eprint
  {http://arxiv.org/abs/1307.2782} {arXiv:1307.2782 [hep-ex]} \BibitemShut
  {NoStop}%
\bibitem [{\citenamefont {Beneke}\ \emph {et~al.}(2005)\citenamefont {Beneke},
  \citenamefont {Feldmann},\ and\ \citenamefont {Seidel}}]{Beneke:2004dp}%
  \BibitemOpen
  \bibfield  {author} {\bibinfo {author} {\bibfnamefont {M.}~\bibnamefont
  {Beneke}}, \bibinfo {author} {\bibfnamefont {Th.}\ \bibnamefont {Feldmann}},
  \ and\ \bibinfo {author} {\bibfnamefont {D.}~\bibnamefont {Seidel}},\
  }\bibfield  {title} {\enquote {\bibinfo {title} {{Exclusive radiative and
  electroweak $b \to d$ and $b \to s$ penguin decays at NLO}},}\ }\href
  {\doibase 10.1140/epjc/s2005-02181-5} {\bibfield  {journal} {\bibinfo
  {journal} {Eur. Phys. J.}\ }\textbf {\bibinfo {volume} {C41}},\ \bibinfo
  {pages} {173--188} (\bibinfo {year} {2005})},\ \Eprint
  {http://arxiv.org/abs/hep-ph/0412400} {arXiv:hep-ph/0412400 [hep-ph]}
  \BibitemShut {NoStop}%
\bibitem [{Note1()}]{Note1}%
  \BibitemOpen
  \bibinfo {note} {We thank Yuming Wang for providing us with the results for
  $B\to K^*\gamma ^*$ in digital form.}\BibitemShut {Stop}%
\bibitem [{\citenamefont {Ciuchini}\ \emph {et~al.}(2015)\citenamefont
  {Ciuchini}, \citenamefont {Fedele}, \citenamefont {Franco}, \citenamefont
  {Mishima}, \citenamefont {Paul}, \citenamefont {Silvestrini},\ and\
  \citenamefont {Valli}}]{Ciuchini:2015qxb}%
  \BibitemOpen
  \bibfield  {author} {\bibinfo {author} {\bibfnamefont {Marco}\ \bibnamefont
  {Ciuchini}}, \bibinfo {author} {\bibfnamefont {Marco}\ \bibnamefont
  {Fedele}}, \bibinfo {author} {\bibfnamefont {Enrico}\ \bibnamefont {Franco}},
  \bibinfo {author} {\bibfnamefont {Satoshi}\ \bibnamefont {Mishima}}, \bibinfo
  {author} {\bibfnamefont {Ayan}\ \bibnamefont {Paul}}, \bibinfo {author}
  {\bibfnamefont {Luca}\ \bibnamefont {Silvestrini}}, \ and\ \bibinfo {author}
  {\bibfnamefont {Mauro}\ \bibnamefont {Valli}},\ }\bibfield  {title} {\enquote
  {\bibinfo {title} {{$B\to K^* \ell^+ \ell^-$ decays at large recoil in the
  Standard Model: a theoretical reappraisal}},}\ }\href@noop {} {\  (\bibinfo
  {year} {2015})},\ \Eprint {http://arxiv.org/abs/1512.07157} {arXiv:1512.07157
  [hep-ph]} \BibitemShut {NoStop}%
\bibitem [{\citenamefont {Descotes-Genon}\ \emph
  {et~al.}(2013{\natexlab{b}})\citenamefont {Descotes-Genon}, \citenamefont
  {Matias}, \citenamefont {Ramon},\ and\ \citenamefont
  {Virto}}]{DescotesGenon:2012zf}%
  \BibitemOpen
  \bibfield  {author} {\bibinfo {author} {\bibfnamefont {Sebastien}\
  \bibnamefont {Descotes-Genon}}, \bibinfo {author} {\bibfnamefont {Joaquim}\
  \bibnamefont {Matias}}, \bibinfo {author} {\bibfnamefont {Marc}\ \bibnamefont
  {Ramon}}, \ and\ \bibinfo {author} {\bibfnamefont {Javier}\ \bibnamefont
  {Virto}},\ }\bibfield  {title} {\enquote {\bibinfo {title} {{Implications
  from clean observables for the binned analysis of $B \to K^*\mu^+\mu^-$ at
  large recoil}},}\ }\href {\doibase 10.1007/JHEP01(2013)048} {\bibfield
  {journal} {\bibinfo  {journal} {JHEP}\ }\textbf {\bibinfo {volume} {01}},\
  \bibinfo {pages} {048} (\bibinfo {year} {2013}{\natexlab{b}})},\ \Eprint
  {http://arxiv.org/abs/1207.2753} {arXiv:1207.2753 [hep-ph]} \BibitemShut
  {NoStop}%
\bibitem [{\citenamefont {Aaij}\ \emph
  {et~al.}(2013{\natexlab{b}})\citenamefont {Aaij} \emph
  {et~al.}}]{Aaij:2013qta}%
  \BibitemOpen
  \bibfield  {author} {\bibinfo {author} {\bibfnamefont {R.}~\bibnamefont
  {Aaij}} \emph {et~al.} (\bibinfo {collaboration} {LHCb}),\ }\bibfield
  {title} {\enquote {\bibinfo {title} {{Measurement of Form-Factor-Independent
  Observables in the Decay $B^{0} \to K^{*0} \mu^+ \mu^-$}},}\ }\href {\doibase
  10.1103/PhysRevLett.111.191801} {\bibfield  {journal} {\bibinfo  {journal}
  {Phys. Rev. Lett.}\ }\textbf {\bibinfo {volume} {111}},\ \bibinfo {pages}
  {191801} (\bibinfo {year} {2013}{\natexlab{b}})},\ \Eprint
  {http://arxiv.org/abs/1308.1707} {arXiv:1308.1707 [hep-ex]} \BibitemShut
  {NoStop}%
\bibitem [{\citenamefont {Amhis}\ \emph {et~al.}(2016)\citenamefont {Amhis}
  \emph {et~al.}}]{Amhis:2016xyh}%
  \BibitemOpen
  \bibfield  {author} {\bibinfo {author} {\bibfnamefont {Y.}~\bibnamefont
  {Amhis}} \emph {et~al.},\ }\bibfield  {title} {\enquote {\bibinfo {title}
  {{Averages of $b$-hadron, $c$-hadron, and $\tau$-lepton properties as of
  summer 2016}},}\ }\href@noop {} {\  (\bibinfo {year} {2016})},\ \Eprint
  {http://arxiv.org/abs/1612.07233} {arXiv:1612.07233 [hep-ex]} \BibitemShut
  {NoStop}%
\bibitem [{\citenamefont {Paul}\ and\ \citenamefont
  {Straub}(2017)}]{Paul:2016urs}%
  \BibitemOpen
  \bibfield  {author} {\bibinfo {author} {\bibfnamefont {Ayan}\ \bibnamefont
  {Paul}}\ and\ \bibinfo {author} {\bibfnamefont {David~M.}\ \bibnamefont
  {Straub}},\ }\bibfield  {title} {\enquote {\bibinfo {title} {{Constraints on
  new physics from radiative $B$ decays}},}\ }\href {\doibase
  10.1007/JHEP04(2017)027} {\bibfield  {journal} {\bibinfo  {journal} {JHEP}\
  }\textbf {\bibinfo {volume} {04}},\ \bibinfo {pages} {027} (\bibinfo {year}
  {2017})},\ \Eprint {http://arxiv.org/abs/1608.02556} {arXiv:1608.02556
  [hep-ph]} \BibitemShut {NoStop}%
\bibitem [{\citenamefont {Altmannshofer}\ \emph {et~al.}(2009)\citenamefont
  {Altmannshofer}, \citenamefont {Ball}, \citenamefont {Bharucha},
  \citenamefont {Buras}, \citenamefont {Straub},\ and\ \citenamefont
  {Wick}}]{Altmannshofer:2008dz}%
  \BibitemOpen
  \bibfield  {author} {\bibinfo {author} {\bibfnamefont {Wolfgang}\
  \bibnamefont {Altmannshofer}}, \bibinfo {author} {\bibfnamefont {Patricia}\
  \bibnamefont {Ball}}, \bibinfo {author} {\bibfnamefont {Aoife}\ \bibnamefont
  {Bharucha}}, \bibinfo {author} {\bibfnamefont {Andrzej~J.}\ \bibnamefont
  {Buras}}, \bibinfo {author} {\bibfnamefont {David~M.}\ \bibnamefont
  {Straub}}, \ and\ \bibinfo {author} {\bibfnamefont {Michael}\ \bibnamefont
  {Wick}},\ }\bibfield  {title} {\enquote {\bibinfo {title} {{Symmetries and
  Asymmetries of $B \to K^{*} \mu^{+} \mu^{-}$ Decays in the Standard Model and
  Beyond}},}\ }\href {\doibase 10.1088/1126-6708/2009/01/019} {\bibfield
  {journal} {\bibinfo  {journal} {JHEP}\ }\textbf {\bibinfo {volume} {01}},\
  \bibinfo {pages} {019} (\bibinfo {year} {2009})},\ \Eprint
  {http://arxiv.org/abs/0811.1214} {arXiv:0811.1214 [hep-ph]} \BibitemShut
  {NoStop}%
\bibitem [{\citenamefont {Aaij}\ \emph
  {et~al.}(2016{\natexlab{a}})\citenamefont {Aaij} \emph
  {et~al.}}]{Aaij:2015oid}%
  \BibitemOpen
  \bibfield  {author} {\bibinfo {author} {\bibfnamefont {Roel}\ \bibnamefont
  {Aaij}} \emph {et~al.} (\bibinfo {collaboration} {LHCb}),\ }\bibfield
  {title} {\enquote {\bibinfo {title} {{Angular analysis of the $B^{0} \to
  K^{*0} \mu^{+} \mu^{-}$ decay using 3 fb$^{-1}$ of integrated luminosity}},}\
  }\href {\doibase 10.1007/JHEP02(2016)104} {\bibfield  {journal} {\bibinfo
  {journal} {JHEP}\ }\textbf {\bibinfo {volume} {02}},\ \bibinfo {pages} {104}
  (\bibinfo {year} {2016}{\natexlab{a}})},\ \Eprint
  {http://arxiv.org/abs/1512.04442} {arXiv:1512.04442 [hep-ex]} \BibitemShut
  {NoStop}%
\bibitem [{\citenamefont {Aaij}\ \emph
  {et~al.}(2016{\natexlab{b}})\citenamefont {Aaij} \emph
  {et~al.}}]{Aaij:2016flj}%
  \BibitemOpen
  \bibfield  {author} {\bibinfo {author} {\bibfnamefont {Roel}\ \bibnamefont
  {Aaij}} \emph {et~al.} (\bibinfo {collaboration} {LHCb}),\ }\bibfield
  {title} {\enquote {\bibinfo {title} {{Measurements of the S-wave fraction in
  $B^{0}\rightarrow K^{+}\pi^{-}\mu^{+}\mu^{-}$ decays and the
  $B^{0}\rightarrow K^{\ast}(892)^{0}\mu^{+}\mu^{-}$ differential branching
  fraction}},}\ }\href {\doibase 10.1007/JHEP11(2016)047,
  10.1007/JHEP04(2017)142, 10.1007/JHEP11(2016)047, 10.1007/JHEP04(2017)142}
  {\bibfield  {journal} {\bibinfo  {journal} {JHEP}\ }\textbf {\bibinfo
  {volume} {11}},\ \bibinfo {pages} {047} (\bibinfo {year}
  {2016}{\natexlab{b}})},\ \Eprint {http://arxiv.org/abs/1606.04731}
  {arXiv:1606.04731 [hep-ex]} \BibitemShut {NoStop}%
\bibitem [{\citenamefont {Beaujean}\ \emph {et~al.}(2015)\citenamefont
  {Beaujean}, \citenamefont {Chrzaszcz}, \citenamefont {Serra},\ and\
  \citenamefont {van Dyk}}]{Beaujean:2015xea}%
  \BibitemOpen
  \bibfield  {author} {\bibinfo {author} {\bibfnamefont {Frederik}\
  \bibnamefont {Beaujean}}, \bibinfo {author} {\bibfnamefont {Marcin}\
  \bibnamefont {Chrzaszcz}}, \bibinfo {author} {\bibfnamefont {Nicola}\
  \bibnamefont {Serra}}, \ and\ \bibinfo {author} {\bibfnamefont {Danny}\
  \bibnamefont {van Dyk}},\ }\bibfield  {title} {\enquote {\bibinfo {title}
  {{Extracting Angular Observables without a Likelihood and Applications to
  Rare Decays}},}\ }\href {\doibase 10.1103/PhysRevD.91.114012} {\bibfield
  {journal} {\bibinfo  {journal} {Phys. Rev.}\ }\textbf {\bibinfo {volume}
  {D91}},\ \bibinfo {pages} {114012} (\bibinfo {year} {2015})},\ \Eprint
  {http://arxiv.org/abs/1503.04100} {arXiv:1503.04100 [hep-ex]} \BibitemShut
  {NoStop}%
\bibitem [{\citenamefont {Beaujean}\ \emph {et~al.}(2014)\citenamefont
  {Beaujean}, \citenamefont {Bobeth},\ and\ \citenamefont {van
  Dyk}}]{Beaujean:2013soa}%
  \BibitemOpen
  \bibfield  {author} {\bibinfo {author} {\bibfnamefont {Frederik}\
  \bibnamefont {Beaujean}}, \bibinfo {author} {\bibfnamefont {Christoph}\
  \bibnamefont {Bobeth}}, \ and\ \bibinfo {author} {\bibfnamefont {Danny}\
  \bibnamefont {van Dyk}},\ }\bibfield  {title} {\enquote {\bibinfo {title}
  {{Comprehensive Bayesian analysis of rare (semi)leptonic and radiative $B$
  decays}},}\ }\href {\doibase 10.1140/epjc/s10052-014-2897-0,
  10.1140/epjc/s10052-014-3179-6} {\bibfield  {journal} {\bibinfo  {journal}
  {Eur. Phys. J.}\ }\textbf {\bibinfo {volume} {C74}},\ \bibinfo {pages} {2897}
  (\bibinfo {year} {2014})},\ \bibinfo {note} {[Erratum: Eur. Phys.
  J.C74,3179(2014)]},\ \Eprint {http://arxiv.org/abs/1310.2478}
  {arXiv:1310.2478 [hep-ph]} \BibitemShut {NoStop}%
\bibitem [{\citenamefont {Descotes-Genon}\ \emph
  {et~al.}(2013{\natexlab{c}})\citenamefont {Descotes-Genon}, \citenamefont
  {Matias},\ and\ \citenamefont {Virto}}]{Descotes-Genon:2013wba}%
  \BibitemOpen
  \bibfield  {author} {\bibinfo {author} {\bibfnamefont {Sebastien}\
  \bibnamefont {Descotes-Genon}}, \bibinfo {author} {\bibfnamefont {Joaquim}\
  \bibnamefont {Matias}}, \ and\ \bibinfo {author} {\bibfnamefont {Javier}\
  \bibnamefont {Virto}},\ }\bibfield  {title} {\enquote {\bibinfo {title}
  {{Understanding the $B\to K^*\mu^+\mu^-$ Anomaly}},}\ }\href {\doibase
  10.1103/PhysRevD.88.074002} {\bibfield  {journal} {\bibinfo  {journal} {Phys.
  Rev.}\ }\textbf {\bibinfo {volume} {D88}},\ \bibinfo {pages} {074002}
  (\bibinfo {year} {2013}{\natexlab{c}})},\ \Eprint
  {http://arxiv.org/abs/1307.5683} {arXiv:1307.5683 [hep-ph]} \BibitemShut
  {NoStop}%
\bibitem [{\citenamefont {Altmannshofer}\ and\ \citenamefont
  {Straub}(2013)}]{Altmannshofer:2013foa}%
  \BibitemOpen
  \bibfield  {author} {\bibinfo {author} {\bibfnamefont {Wolfgang}\
  \bibnamefont {Altmannshofer}}\ and\ \bibinfo {author} {\bibfnamefont
  {David~M.}\ \bibnamefont {Straub}},\ }\bibfield  {title} {\enquote {\bibinfo
  {title} {{New Physics in $B \to K^*\mu\mu$?}}}\ }\href {\doibase
  10.1140/epjc/s10052-013-2646-9} {\bibfield  {journal} {\bibinfo  {journal}
  {Eur. Phys. J.}\ }\textbf {\bibinfo {volume} {C73}},\ \bibinfo {pages} {2646}
  (\bibinfo {year} {2013})},\ \Eprint {http://arxiv.org/abs/1308.1501}
  {arXiv:1308.1501 [hep-ph]} \BibitemShut {NoStop}%
\bibitem [{\citenamefont {Altmannshofer}\ \emph {et~al.}(2017)\citenamefont
  {Altmannshofer}, \citenamefont {Niehoff}, \citenamefont {Stangl},\ and\
  \citenamefont {Straub}}]{Altmannshofer:2017fio}%
  \BibitemOpen
  \bibfield  {author} {\bibinfo {author} {\bibfnamefont {Wolfgang}\
  \bibnamefont {Altmannshofer}}, \bibinfo {author} {\bibfnamefont {Christoph}\
  \bibnamefont {Niehoff}}, \bibinfo {author} {\bibfnamefont {Peter}\
  \bibnamefont {Stangl}}, \ and\ \bibinfo {author} {\bibfnamefont {David~M.}\
  \bibnamefont {Straub}},\ }\bibfield  {title} {\enquote {\bibinfo {title}
  {{Status of the $B\rightarrow K^*\mu ^+\mu ^-$ anomaly after Moriond
  2017}},}\ }\href {\doibase 10.1140/epjc/s10052-017-4952-0} {\bibfield
  {journal} {\bibinfo  {journal} {Eur. Phys. J.}\ }\textbf {\bibinfo {volume}
  {C77}},\ \bibinfo {pages} {377} (\bibinfo {year} {2017})},\ \Eprint
  {http://arxiv.org/abs/1703.09189} {arXiv:1703.09189 [hep-ph]} \BibitemShut
  {NoStop}%
\bibitem [{\citenamefont {Capdevila}\ \emph {et~al.}(2017)\citenamefont
  {Capdevila}, \citenamefont {Crivellin}, \citenamefont {Descotes-Genon},
  \citenamefont {Matias},\ and\ \citenamefont {Virto}}]{Capdevila:2017bsm}%
  \BibitemOpen
  \bibfield  {author} {\bibinfo {author} {\bibfnamefont {Bernat}\ \bibnamefont
  {Capdevila}}, \bibinfo {author} {\bibfnamefont {Andreas}\ \bibnamefont
  {Crivellin}}, \bibinfo {author} {\bibfnamefont {S\'ebastien}\ \bibnamefont
  {Descotes-Genon}}, \bibinfo {author} {\bibfnamefont {Joaquim}\ \bibnamefont
  {Matias}}, \ and\ \bibinfo {author} {\bibfnamefont {Javier}\ \bibnamefont
  {Virto}},\ }\bibfield  {title} {\enquote {\bibinfo {title} {{Patterns of New
  Physics in $b\to s\ell^+\ell^-$ transitions in the light of recent data}},}\
  }\href@noop {} {\  (\bibinfo {year} {2017})},\ \Eprint
  {http://arxiv.org/abs/1704.05340} {arXiv:1704.05340 [hep-ph]} \BibitemShut
  {NoStop}%
\bibitem [{\citenamefont {Geng}\ \emph {et~al.}(2017)\citenamefont {Geng},
  \citenamefont {Grinstein}, \citenamefont {J{\"a}ger}, \citenamefont
  {Martin~Camalich}, \citenamefont {Ren},\ and\ \citenamefont
  {Shi}}]{Geng:2017svp}%
  \BibitemOpen
  \bibfield  {author} {\bibinfo {author} {\bibfnamefont {Li-Sheng}\
  \bibnamefont {Geng}}, \bibinfo {author} {\bibfnamefont {Benjamín}\
  \bibnamefont {Grinstein}}, \bibinfo {author} {\bibfnamefont {Sebastian}\
  \bibnamefont {J{\"a}ger}}, \bibinfo {author} {\bibfnamefont {Jorge}\
  \bibnamefont {Martin~Camalich}}, \bibinfo {author} {\bibfnamefont {Xiu-Lei}\
  \bibnamefont {Ren}}, \ and\ \bibinfo {author} {\bibfnamefont {Rui-Xiang}\
  \bibnamefont {Shi}},\ }\bibfield  {title} {\enquote {\bibinfo {title}
  {{Towards the discovery of new physics with lepton-universality ratios of
  $b\to s\ell\ell$ decays}},}\ }\href@noop {} {\  (\bibinfo {year} {2017})},\
  \Eprint {http://arxiv.org/abs/1704.05446} {arXiv:1704.05446 [hep-ph]}
  \BibitemShut {NoStop}%
\bibitem [{\citenamefont {Hurth}\ \emph {et~al.}(2017)\citenamefont {Hurth},
  \citenamefont {Mahmoudi}, \citenamefont {Martinez~Santos},\ and\
  \citenamefont {Neshatpour}}]{Hurth:2017hxg}%
  \BibitemOpen
  \bibfield  {author} {\bibinfo {author} {\bibfnamefont {T.}~\bibnamefont
  {Hurth}}, \bibinfo {author} {\bibfnamefont {F.}~\bibnamefont {Mahmoudi}},
  \bibinfo {author} {\bibfnamefont {D.}~\bibnamefont {Martinez~Santos}}, \ and\
  \bibinfo {author} {\bibfnamefont {S.}~\bibnamefont {Neshatpour}},\ }\bibfield
   {title} {\enquote {\bibinfo {title} {{Update on lepton non-universality in
  exclusive $b\to s \ell\ell$ decays}},}\ }\href@noop {} {\  (\bibinfo {year}
  {2017})},\ \Eprint {http://arxiv.org/abs/1705.06274} {arXiv:1705.06274
  [hep-ph]} \BibitemShut {NoStop}%
\bibitem [{\citenamefont {Bona}\ \emph {et~al.}(2006)\citenamefont {Bona} \emph
  {et~al.}}]{Bona:2006ah}%
  \BibitemOpen
  \bibfield  {author} {\bibinfo {author} {\bibfnamefont {M.}~\bibnamefont
  {Bona}} \emph {et~al.} (\bibinfo {collaboration} {UTfit Collaboration}),\
  }\bibfield  {title} {\enquote {\bibinfo {title} {{The Unitarity Triangle Fit
  in the Standard Model and Hadronic Parameters from Lattice QCD: A Reappraisal
  after the Measurements of $\Delta m_s$ and BR($B \to \tau \nu_\tau$)}},}\
  }\href {\doibase 10.1088/1126-6708/2006/10/081} {\bibfield  {journal}
  {\bibinfo  {journal} {JHEP}\ }\textbf {\bibinfo {volume} {0610}},\ \bibinfo
  {pages} {081} (\bibinfo {year} {2006})},\ \bibinfo {note} {we use the updated
  data from Winter 2013 (pre-Moriond 13)},\ \Eprint
  {http://arxiv.org/abs/hep-ph/0606167} {arXiv:hep-ph/0606167 [hep-ph]}
  \BibitemShut {NoStop}%
\bibitem [{\citenamefont {Charles}\ \emph {et~al.}(1999)\citenamefont
  {Charles}, \citenamefont {Le~Yaouanc}, \citenamefont {Oliver}, \citenamefont
  {Pene},\ and\ \citenamefont {Raynal}}]{Charles:1998dr}%
  \BibitemOpen
  \bibfield  {author} {\bibinfo {author} {\bibfnamefont {J.}~\bibnamefont
  {Charles}}, \bibinfo {author} {\bibfnamefont {A.}~\bibnamefont {Le~Yaouanc}},
  \bibinfo {author} {\bibfnamefont {L.}~\bibnamefont {Oliver}}, \bibinfo
  {author} {\bibfnamefont {O.}~\bibnamefont {Pene}}, \ and\ \bibinfo {author}
  {\bibfnamefont {J.~C.}\ \bibnamefont {Raynal}},\ }\bibfield  {title}
  {\enquote {\bibinfo {title} {{Heavy to light form-factors in the heavy mass
  to large energy limit of QCD}},}\ }\href {\doibase
  10.1103/PhysRevD.60.014001} {\bibfield  {journal} {\bibinfo  {journal} {Phys.
  Rev.}\ }\textbf {\bibinfo {volume} {D60}},\ \bibinfo {pages} {014001}
  (\bibinfo {year} {1999})},\ \Eprint {http://arxiv.org/abs/hep-ph/9812358}
  {arXiv:hep-ph/9812358 [hep-ph]} \BibitemShut {NoStop}%
\bibitem [{\citenamefont {Beneke}\ and\ \citenamefont
  {Feldmann}(2001)}]{Beneke:2000wa}%
  \BibitemOpen
  \bibfield  {author} {\bibinfo {author} {\bibfnamefont {M.}~\bibnamefont
  {Beneke}}\ and\ \bibinfo {author} {\bibfnamefont {T.}~\bibnamefont
  {Feldmann}},\ }\bibfield  {title} {\enquote {\bibinfo {title} {{Symmetry
  breaking corrections to heavy to light B meson form-factors at large
  recoil}},}\ }\href {\doibase 10.1016/S0550-3213(00)00585-X} {\bibfield
  {journal} {\bibinfo  {journal} {Nucl. Phys.}\ }\textbf {\bibinfo {volume}
  {B592}},\ \bibinfo {pages} {3--34} (\bibinfo {year} {2001})},\ \Eprint
  {http://arxiv.org/abs/hep-ph/0008255} {arXiv:hep-ph/0008255 [hep-ph]}
  \BibitemShut {NoStop}%
\end{thebibliography}%


\end{document}